\documentclass[iop,apj,twocolappendix]{emulateapj}
\usepackage[varg]{txfonts}
\usepackage{bm}

\newcommand{\beq}{\begin{equation}}
\newcommand{\eeq}{\end{equation}}
\newcommand{\beqn}{\begin{eqnarray}}
\newcommand{\eeqn}{\end{eqnarray}}

\newcommand{\AU}{{\rm AU}}

\newcommand{\brac}[1]{\langle #1 \rangle}
\newcommand{\eqref}[1]{(\ref{#1})}

\newcommand{\dfrac}[2]{ {\displaystyle\frac{#1}{#2}} }
\newcommand{\pfrac}[2]{ \biggl(\dfrac{#1}{#2}\biggr) }

\newcommand{\md}{{\rm mid}}
\newcommand{\ideal}{{\rm ideal}}
\newcommand{\res}{{\rm res}}

\defcitealias{O09}{O09}
\defcitealias{OH11}{OH11}
\defcitealias{GNT12}{GNT12}
\defcitealias{OO13b}{Paper II}

\slugcomment{ApJ accepted}

\shorttitle{The fate of planetesimals in turbulent disks. I}
\shortauthors{Okuzumi \& Ormel}

\begin{document}

\title{The Fate of Planetesimals in Turbulent Disks with Dead Zones. \\I. The Turbulent Stirring Recipe}
\author{Satoshi Okuzumi\altaffilmark{1,2,3} and Chris W. Ormel\altaffilmark{4,5}}
\altaffiltext{1}{Department of Earth and Planetary Sciences, 
Tokyo Institute of Technology, Meguro-ku, Tokyo, 152-8551; okuzumi@geo.titech.ac.jp}
\altaffiltext{2}{Department of Physics, Nagoya University, Nagoya, Aichi 464-8602, Japan}
\altaffiltext{3}{JSPS Superlative Research Fellow}
\altaffiltext{4}{Astronomy Department, University of California, Berkeley, CA 94720, USA}
\altaffiltext{5}{Hubble Fellow}

\begin{abstract}
Turbulence in protoplanetary disks affects planet formation in many ways. While small dust particles are mainly affected by the aerodynamical coupling with turbulent gas velocity fields, planetesimals and larger bodies are more affected by gravitational interaction with gas density fluctuations. For the latter process, a number of numerical simulations have been performed in recent years, but a fully parameter-independent understanding has not been yet established. In this study, we present simple scaling relations for the planetesimal stirring rate in turbulence driven by magnetorotational instability (MRI), taking into account the stabilization of MRI due to Ohmic resistivity. We begin with order-of-magnitude estimates of the turbulence-induced gravitational force acting on solid bodies and associated diffusion coefficients for their orbital elements. We then test the predicted scaling relations using the results of recent Ohmic-resistive MHD simulations by Gressel et al. We find that these relations successfully explain the simulation results if we properly fix order-of-unity uncertainties within the estimates. We also update the saturation predictor for the density fluctuation amplitude in MRI-driven turbulence originally proposed by Okuzumi \& Hirose. Combination of the scaling relations and saturation predictor allows to know how the turbulent stirring rate of planetesimals depends on disk parameters such as the gas column density, distance from the central star, vertical resistivity distribution, and net vertical magnetic flux. In Paper II, we apply our recipe to planetesimal accretion to discuss its viability in turbulent disks.
\end{abstract}

\maketitle

\section{Introduction}
Planets are believed to form in circumstellar gas disks called protoplanetary disks.
Planet formation begins with coagulation of submicron-sized dust particles through intermolecular forces.
This stage is followed by the formation of kilometer-sized planetesimals through 
the gravitational collapse of microscopic dust aggregates mediated by gravitational \citep{GW73} or streaming \citep{YG05,J+07} instabilities, or through 
further dust coagulation \citep[e.g.,][]{OTKW12,W+12a}.
Large planetesimals experience runaway growth mediated by gravitational focusing \citep{WS89,KI96}, forming even larger solid bodies called protoplanets (or planetary embryos).
In the final stage, protoplanets evolve into gas giants by accreting the disk gas 
or into larger solid planets through giant impacts during/after the dispersal of the gas disk.

The fate of these formation processes crucially depends 
on the turbulent state of the gas disk. 
Turbulence induces a random motion of solid particles smaller than planetesimals 
through the aerodynamical friction force \citep[e.g.,][]{V+80,OC07}. 
The resulting turbulent diffusion acts against accumulation of the solid particles 
\citep{CSP05,TWBY06,FP06,JKM06}, which limits planetesimal formation 
via gravitational instability \citep{Y11}.
The enhanced collision velocity may cause catastrophic disruption of the solid bodies, 
which inhibits direct collisional formation of planetesimals \citep{J+08,OH12}.
Turbulence also accumulates solid particles of particular sizes \citep[e.g.,][]{C+01,J+07},
but its relevance to planetesimal formation is under debate \citep{P+11}.

For planetesimals and larger solid bodies, stochastic gravitational forces 
induced by gas density fluctuations play a more important role.
In turbulent disks, vorticity and nonlinear stress excite gas density fluctuations \citep{HP09a,HP09b}, which give rise to stochastic gravitational forces that 
act on solid bodies.
This particularly affects the motion of large solid bodies that are
well decoupled from the gas friction force, causing  
stochastic orbital migration \citep{LSA04,NP04,N05,JGM06,OMM07,R12} and 
eccentricity stirring \citep{N05,OIM07,IGM08}.

The turbulence-induced eccentricity stirring 
severely constrains the formation of protoplanets at a fundamental level. 
In order for gravitational runaway growth to set in, 
the velocity dispersion of planetesimals must be smaller than their escape velocity \citep{WS89}. 
However, in a fully turbulent disk, this requirement is unlikely to be satisfied 
for planetesimals smaller than 100~km in size \citep{IGM08,NG10}. 
This indicates that runaway growth could be significantly delayed 
depending on the turbulent state of the disks \citep{N05,ODS10b}.
Moreover, the high turbulence-driven relative velocity can make a collision between planetesimals disruptive rather than accumulative, especially in outer regions of the disks  \citep{N05,IGM08,NG10,YMM09,YMM12}.
Thus, to understand the fate of planetesimal growth and succeeding planet formation, 
it is essential to know how gas turbulence is driven in protoplanetary disks, 
and how its strength depends on the disk environment.

One mechanism that can drive strong turbulence 
is the magnetorotational instability \citep[MRI;][]{BH91}.
This is an MHD instability resulting from the coupling 
between a differentially rotating gas disk and magnetic fields.
In an ideal case where the coupling is strong enough, 
the MRI drives strong gas turbulence with the Shakura--Sunyaev parameter 
$\alpha \sim 10^{-2}$ or even larger depending on the strength of 
the net vertical magnetic fields \citep[e.g.,][]{DSP10,SMI10}. 
However, because the ionization degree of protoplanetary disks is 
generally very low, 
non-ideal MHD effects strongly affect the actual level of the turbulence.
For example, a high Ohmic resistivity near the disk midplane 
prevents the coupling between the gas and magnetic fields 
and thereby creates a ``dead zone'' where MRI is inactive \citep{G96}.
The size of the dead zone depends on the ionization degree of the disk gas, 
and is generally large when tiny dust particles that efficiently capture ionized gas particles 
are abundant \citep[e.g.,][]{SMUN00,IN06a}.
Ambipolar diffusion has a similar effect on MRI, but at higher altitudes 
where the gas density is low \citep{B11a,PC11a,PC11b,MET13,DTHK13}. 

Recently, \citet{GNT11} first studied the effect of an Ohmic dead zone on 
turbulent planetesimal stirring.
They performed local stratified MHD simulations at 5~AU taking into account 
a high Ohmic resistivity provided by abundant small dust particles.
They showed that the resulting large dead zone considerably suppresses 
the planetesimal stirring rate.
The effect has been more extensively studied in their latest paper 
(\citealt*{GNT12}; henceforth \citetalias{GNT12}) 
for various values of the net vertical magnetic flux.
They concluded that planetesimal growth beyond the disruption barrier 
is possible in a dead zone if the net flux is so weak that 
upper MRI-active layers do not generate strong density waves. 
This indicates that a dead zone can provide a safe haven for planetesimals.

However, there still remain two open issues. 
First, how much dust is needed to maintain a large enough dead zone?
\citet{GNT11,GNT12} fixed the amount of $0.1~\micron$-sized dust particles 
to be $10~\%$ in mass of the total solids in the disk. 
However, it is unclear whether this amount is reasonable in late stages 
of planet formation where a significant fraction of solids in the disk should 
have been incorporated into planetesimals.
In principle, tiny particles can be resupplied when planetesimals 
undergo collisional fragmentation or erosion.
However, such tiny particles are usually removed immediately 
through their mutual sticking and/or sweep up by larger dust particles.
Thus, the amount of residual dust is determined by the balance between
these competing processes, and therefore cannot be determined {\it a priori}.
Second, can a dead zone act as a safe haven at every location in protoplanetary disks?
The results of \citet{GNT11,GNT12} only apply to 5~AU from the central star, 
but turbulent planetesimal stirring is generally more effective further 
out in disks \citep{IGM08}.
In order to study whether the dead zone is beneficial 
for planetesimal growth in general circumstances, 
a model that does not rely on a specific choice of disk parameters is desirable.

The aim of this study is to provide a general recipe for planetesimal stirring 
in MRI-driven turbulence. 
We construct scaling relations that clarify how the turbulent 
quantities relevant to planetesimal stirring depend on each other and on 
basic disk parameters.
This is an extension of recent work by \citet[][henceforth \citetalias{OH11}]{OH11}.
They performed a systematic set of local stratified MHD simulations 
with a dead zone, and provided an analytic prescription for the amplitude of the gas density fluctuations as a function of the net vertical flux, vertical resistivity profile, and other disk parameters. 
In this paper, we begin with an order-of-magnitude estimate 
to derive scaling relations that link the density fluctuation
amplitude to the turbulent stirring rate of solid bodies.
We then calibrate them using the published data by \citetalias{GNT12}.
We also update the density fluctuation recipe of \citetalias{OH11} 
using the same published data.
An application of our recipe to runaway planetesimal growth will be presented in Paper II \citep{OO13b}.

The plan of this paper is as follows.
In Section~\ref{sec:estimate}, we present order-of-magnitude estimates 
that predict relationships among the density fluctuation, random gravity, and 
orbital diffusion coefficients for planetesimals.
In Section~\ref{sec:calibration}, we compare our predictions with the simulation results presented by \citetalias{GNT12} to present calibrated prescriptions 
for planetesimal stirring.
The predictor function for the density fluctuations is presented 
in Section~\ref{sec:predictor}.
Comparison with previous results relying on ideal MHD and 
implication for planetesimal stirring in protoplanetary disks is given 
in Section~\ref{sec:discussion}.
A summary of this study is given in Section~\ref{sec:summary}.
 
\section{Order-of-Magnitude Estimates}\label{sec:estimate}
{ In order to clarify how the turbulent stirring rate of planetesimals 
generally depends on disk parameters, we begin with deriving 
scaling relations between relevant turbulent quantities from order-of-magnitude arguments.
Verification and calibration of the derived relations will be done in Section~\ref{sec:calibration}.
}

Our estimation follows two steps. 
First, we relate gas density fluctuations to random gravitational forces
on planetesimals using Gauss's law for gravity.
We then relate the random gravity to the diffusion coefficients for planetesimals.
The second step is based on recent work by \citet{RP09} that regards 
the equation of motion for planetesimals as a stochastic differential (or Langevin) equation.

\subsection{Random Gravity}\label{sec:gravity}
\begin{figure*}[t]
\epsscale{0.75}
\plotone{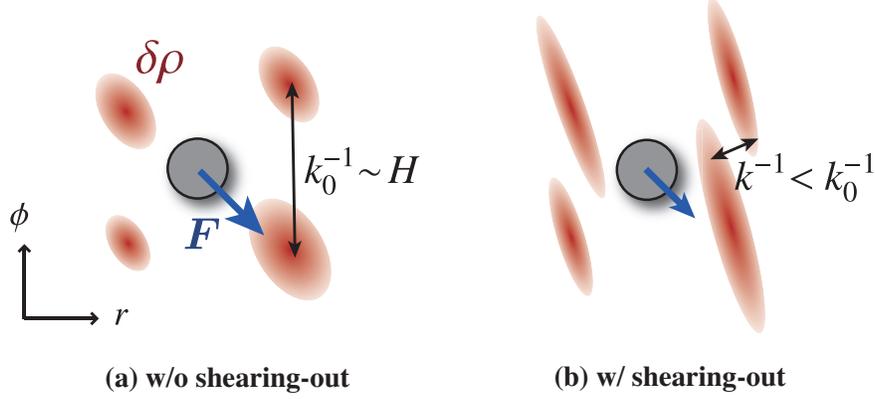}
\caption{Schematic illustration showing how density fluctuations 
create fluctuating gravity fields. The patches indicate density bumps produced by
turbulence, while the circle at the center indicates { a solid body gravitationally interacting with the density fluctuations}.
In MRI-active regions, the characteristic wavenumber $k_0$ of the density fluctuations are of the order $\sim H$ (panel (a)). 
In dead zones, density fluctuations have a higher wavenumber 
because they get sheared out by the background flow when they propagate 
from active regions (panel (b)).}
\label{fig:grav}
\end{figure*}

We denote the gas density perturbation by $\delta\rho$ and 
the induced gravitational force on planetesimals (per unit mass) by ${\bm F}$ 
(see Figure~\ref{fig:grav} (a)).
These are assumed to be stochastic variables
with vanishing mean values $\brac{\delta\rho} = \brac{{\bm F}} = 0$
and nonzero mean square values $\brac{\delta\rho^2}$ and $\brac{{\bm F}^2} \equiv \brac{F^2}$.
The density perturbation and induced gravitational force  
are related to each other by Gauss's law for gravity,
\beq
\nabla\cdot{\bm F} = -4\pi G \delta \rho, 
\label{eq:Gauss}
\eeq
where $G$ is the gravitational constant.

We want to estimate the amplitude of the random gravity for given 
density fluctuation amplitude. This can be done by assuming that 
the density fluctuations have a characteristic wavenumber ${\bm k}$.
With this assumption, we can estimate that  
$|\nabla\cdot{\bm F}| \sim k \brac{F^2}^{1/2}$, where $k = |{\bm k}|$
is the magnitude of ${\bm k}$.
Thus, from Equation~\eqref{eq:Gauss}, we have 
\beq
\brac{F^2}^{1/2} \sim \frac{4\pi G}{k} \brac{\delta\rho^2}^{1/2}.
\label{eq:F}
\eeq

If the disk is entirely MRI-turbulent, the characteristic wavenumber  
(or the inverse of the correlation length) $k$ is of the order 
$\sim 1/H$, where $H$ is the gas scale height
\citep{GGSJ09,HP09a,HP09b,NG10}.
This suggests that 
\beq
\brac{F^2}^{1/2} = {\cal A}_1 GH \brac{\delta\rho^2}^{1/2},
\label{eq:F1}
\eeq
where ${\cal A}_1$ is an order-of-unity number that represents the overall uncertainty in the above estimate.

However, if there is a dead zone at the midplane, we need to take into 
account the shearing-out of the density fluctuations.
In the presence of a dead zone, the sources of the density fluctuation 
at the midplane are density waves that have 
propagated from the upper MRI-active layers. 
At the midplane, these waves have a higher $k$ than they had in the active layers because the differentially rotating background flow shears them out during their propagation (see Figure~\ref{fig:grav} (b)). 
The importance of the shearing-out has first been pointed out by \citetalias{GNT12} { (see their Section~5.1)}, and we will quantify this with the following argument.
Let us denote the radial and azimuthal wavenumbers 
of a density wave by $k_r$ and $k_\phi$, respectively. 
The shearing motion of the gas disks changes the radial wavenumber, 
and this can be expressed as \citep{GL65}
\beq
k_r = k_{r0} + \frac{3}{2}k_{\phi}\Omega \delta t_{\rm travel}.
\label{eq:kr}
\eeq 
where $\Omega$ is the Keplerian frequency, $\delta t_{\rm travel}$ is the time 
passed after the wave is generated, and $k_{r0}$ is the initial value of $k_r$.
We now assume that a dead zone has a vertical extent $|z|\leq H_{\rm DZ}$,
where $H_{\rm DZ}$ is the dead-zone half width.
Then, for density waves at the midplane, 
$\delta t_{\rm travel}$ should be comparable 
to the time the waves travel from the active layer 
to the midplane, i.e., $\delta t_{\rm travel} \sim H_{\rm DZ}/c_s$, 
where $c_s$ is the sound speed.  
This can be rewritten as $\Omega \delta t_{\rm travel} \sim H_{\rm DZ}/H$
since $H = c_s/\Omega$.
Substituting this into Equation~\eqref{eq:kr} and assuming 
$k_{r0} \sim k_{\phi} \sim 1/H$ and $k \sim k_r$, 
the characteristic wavenumber $k$ of the 
density waves at the midplane can be estimated as 
\beq
k \sim  \frac{1}{H}\left( 1+{\cal A}_2 \frac{H_{\rm DZ}}{H}\right),
\eeq
where  ${\cal A}_2$ is another order-of-unity constant.
{ The assumption $k_{r0}\sim k_\phi$ may be somewhat 
inaccurate given the anisotropy of MRI-driven turbulence, 
but its effect is absorbed in ${\cal A}_2$.}
If we use this in Equation~\eqref{eq:F}
we obtain 
\beq
\brac{F^2}^{1/2} = \frac{{\cal A}_1 GH}{1+ {\cal A}_2 H_{\rm DZ}/H}\brac{\delta\rho^2}^{1/2}.
\label{eq:F2}
\eeq
Equation~\eqref{eq:F2} also applies to ideal MRI-turbulent disks
because it reduces to Equation~\eqref{eq:F1} 
in the limit of $H_{\rm DZ} \to 0$.

Equation~\eqref{eq:F2} indicates that in the presence of a dead zone
no simple linear scaling applies to the relation 
between the magnitudes of the random gravity fields and density fluctuations, 
as observed by \citetalias{GNT12}. 
A dead zone reduces the random forcing in two ways: 
directly by suppressing $\brac{\delta\rho^2}^{1/2}$, and 
indirectly by enhancing $k$ (shearing-out).
The factor $(1+{\cal A}_2H_{\rm DZ}/H)^{-1}$
appearing in Equation~\eqref{eq:F2} expresses the second effect.

\subsection{Diffusion Coefficients}\label{sec:diffusion}
In fluctuating gravity fields, the motion of solid bodies can 
be described as a random walk in phase space.
\citet{RP09} formulated this process by treating the equation of motion
for the bodies as a Langevin equation with stochastic forcing.
To describe the results of \citet{RP09}, 
we denote the changes in the semi-major axis $a$ and eccentricity $e$ of 
a body during a time interval $\Delta t$ 
as $\Delta a$ and $\Delta e$, respectively.
A random walk in phase space means that 
the ensemble averages $\brac{(\Delta a)^2}$ and $\brac{(\Delta e)^2}$ grow 
linearly with $\Delta t$ on a timescale much longer than 
the correlation time $\tau_c$ of the fluctuating gravity.
This process can be characterized by constant diffusion coefficients
\beq
D_a \equiv \frac{1}{2} \frac{\brac{(\Delta a)^2}}{\Delta t},
\label{eq:Da_def}
\eeq
\beq
D_e \equiv \frac{1}{2} \frac{\brac{(\Delta e)^2}}{\Delta t}.
\label{eq:De_def}
\eeq

\citet{RP09} derived the exact expressions of 
$\brac{(\Delta a)^2}$ and $\brac{(\Delta e)^2}$ under stochastic force ${\bm F}$.
These read
\beq
\brac{(\Delta a)^2} = \frac{8 \brac{F_\phi^2} \tau_c}{\Omega^2} \Delta t,
\label{eq:da2_RP09}
\eeq
\beq
\brac{(\Delta e)^2} = 
\frac{(\brac{F_r^2} +4\brac{F_rF_\phi} + 4\brac{F_\phi^2}) \tau_c}{(1+\Omega^2\tau_c^2)a^2\Omega^2} \Delta t,
\label{eq:de2_RP09}
\eeq
where $\brac{F_r^2}$ and $\brac{F_\phi^2}$ are the mean squared amplitudes 
of the radial and azimuthal components of ${\bm F}$, respectively 
(see Equations~(46) and (47) of \citealt*{RP09}\footnote{Note that 
Equations~(46) and (47) of \citet{RP09} implicitly assume $\brac{F_r^2}=\brac{F_\phi^2}$ and $\brac{F_rF_\phi} = 0$ 
while our Equation~\eqref{eq:da2_RP09} and \eqref{eq:de2_RP09} allow
$\brac{F_r^2} \not= \brac{F_\phi^2}$ and $\brac{F_rF_\phi} \not= 0$.}).  
The corresponding diffusion coefficients are
\beq
D_a 
= \frac{4 \brac{F_\phi^2} \tau_c}{\Omega^2},
\eeq
\beq
D_e 
= \frac{(\brac{F_r^2} + 4\brac{F_rF_\phi} + 4\brac{F_\phi^2}) \tau_c}{2(1+\Omega^2\tau_c^2)a^2\Omega^2}. 
\eeq

For MRI-driven turbulence, $\tau_c \sim \Omega^{-1}$ is suggested by
a number of simulations \citep[e.g.,][]{SITS04,GNT11,GNT12}.
It is also likely that $\brac{F_r^2} \sim  \brac{F_rF_\phi} 
\sim \brac{F_\phi^2}$ within an accuracy of order unity.
Therefore, we anticipate that the diffusion coefficients are of the forms 
\beq
D_a = \frac{{\cal A}_a}{\Omega^3}\brac{F_\phi^2},
\label{eq:Da}
\eeq
\beq
D_e = \frac{{\cal A}_e}{a^2\Omega^3}\brac{F_\phi^2},
\label{eq:De}
\eeq
where ${\cal A}_a$ and ${\cal A}_e$ are order-of-unity constants.
{ Note that the uncertainties about the anisotropy of the random force, 
$\brac{F_r^2}/\brac{F_rF_\phi}$ and $\brac{F_r^2}/\brac{F_\phi^2}$, 
are absorbed in these constants.}

\section{Calibration with GNT12 Data}\label{sec:calibration}
In the previous section, we have predicted how turbulent quantities 
relevant to the orbital evolution of planetesimals 
should be related to each other.
Here, we test these predictions using the published data 
of MHD simulations by \citetalias{GNT12}. 
Our goal is to determine the order-of-unity constants 
involved in the predicted relationships (${\cal A}_1$, ${\cal A}_2$, ${\cal A}_a$, 
and ${\cal A}_e$; see Equations~\eqref{eq:F2}, \eqref{eq:Da}, and \eqref{eq:De}). 

\citetalias{GNT12} conducted local stratified resistive MHD simulations
at 5~AU from the central star 
with different sets of the disk mass, ionization strength, and net vertical field strength.
The gas surface density  was given by $\Sigma = f_\Sigma \times 135~{\rm g~cm^{-2}}$
with $f_\Sigma$ (= 1, 2, or 4) being a dimensionless factor.
The stellar mass and disk aspect ratio are fixed to $M_* = M_\odot$
and $H/a = 0.05$, respectively. 
The Ohmic resistivity was calculated from the balance between 
external ionization and recombination in the gas phase and on dust grains.
The ionization rate $\zeta$ was the sum of the contributions from cosmic rays
 ($\zeta_{\rm CR}$), X-rays ($\zeta_{\rm XR}$), 
and short-lived radionuclides ($\zeta_{\rm SR}$), with $\zeta_{\rm XR}$ and 
$\zeta_{\rm SR}$ being chosen as 10 and $f_{\rm XR}$ (= 1 or 20) times the standard values, respectively.
The dust-to-gas mass ratio was fixed to $10^{-3}$, and the size of 
the dust grains was chosen to be $0.1~\micron$.
The results of all these simulations are summarized in Table 3 of \citetalias{GNT12}.
We will use these data to test and calibrate our order-of-magnitude relations.

One of the most important parameter is the net vertical magnetic flux $\brac{B_z}$. 
This is a conserved quantity in a local-box simulation, 
and determines the strength of MRI turbulence in the saturated state
\citep[e.g.,][]{HGB95,SITS04,SMI10,OH11}. 
In the \citetalias{GNT12} simulations, 
$\brac{B_z}$ was chosen in the range 2.68--46 mG.
\citetalias{GNT12} also conducted a run with a higher net flux ($\brac{B_z} = 86~{\rm mG}$), but the run resulted in an essentially laminar final state with no sustained MRI turbulence.

Table~\ref{tab:param} lists the values of $f_\Sigma$, $f_{\rm XR}$ 
and $\brac{B_z}$ for all simulations presented by \citetalias{GNT12}.
Run A1 and B1 are ideal MHD simulations while runs labeled 
by `D' include Ohmic diffusion.

\begin{deluxetable*}{lcccccccccccc}
\tabletypesize{\scriptsize}
\tablecaption{Model Parameters and Key Observed Quantities of the \citetalias{GNT12} Simulations}
\tablecolumns{13}
\tablewidth{0pt}
\tablehead{
\colhead{Run} & \colhead{$f_\Sigma$} &  \colhead{$f_{\rm XR}$} &
\colhead{$\brac{B_z}$}  & \colhead{$\beta_{z0}$} &
\colhead{$H_{\ideal,0}$} & \colhead{$H_{\Lambda,0}$}  &  \colhead{$H_{\res,0}$} 
& \colhead{$H_{\ideal,\infty}$} 
& \colhead{$\brac{\delta\rho^2}^{1/2}_\md$} 
& \colhead{$\brac{F_\phi^2}^{1/2}$} 
& \colhead{$D_a$}  
& \colhead{$D_e$} 
  \\[1mm]
 & & &  \colhead{$({\rm mG})$} &  & \colhead{$(H)$} & \colhead{$(H)$} & \colhead{$(H)$} & \colhead{$(H)$} &  
\colhead{$(10^{-4}\Omega^2/G)$} &  \colhead{$(10^{-4}H\Omega^2)$} 
& \colhead{$(10^{-7}H^2\Omega)$} &  \colhead{$(10^{-7}H^2\Omega/a^2)$}  
}
\startdata
A1\tablenotemark{a}& 1 & \nodata & 10.7 & $1.4\times10^4$ & 3.2 & 0 & 0 & 1.8 & \nodata & 5.0 & 22. & 13. \\
B1\tablenotemark{a} & 1 & \nodata & 16.1 & $6.3\times10^3$ & 3.0 & 0 & 0 & 1.3 & 5.1 & 8.3 & 47. & 14. \\
D2 & 1 & 20 & 10.7 & $1.4\times10^4$ & 3.2 & 1.6 & 1.2 & 1.8 & 5.0 & 1.5 & 0.42 & 0.12 \\
D1 & 1 & 1 & 10.7 & $1.4\times10^4$ & 3.2 & 2.2 & 1.8 & 1.8 & 3.1 & 0.67 & 0.18 & 0.044 \\
D1.1 & 1 & 1 & 10.7 & $1.4\times10^4$ & 3.2 & 2.2 & 1.8 & 1.8 & 2.8 & 0.60 & 0.17 & 0.052 \\
D1.2 & 2 & 1 & 10.7 & $2.8\times10^4$ & 3.4 & 2.5 & 2.1 & 2.2 & 3.5 & 0.58 & 0.15 & 0.024 \\
D1.4 & 4 & 1 & 10.7 & $5.7\times10^4$ & 3.6 & 2.7 & 2.4 & 2.5 & 4.6 & 0.58 & 0.071 & 0.030 \\
D1.4b & 4 & 1 & 5.37 & $2.3\times10^5$ & 4.0 & 2.9 & 2.6 & 3.0 & 2.5 & 0.31 & 0.024 & 0.0083 \\
D1-WF & 1 & 1 & 2.68 & $2.3\times10^5$ & 4.0 & 2.7 & 2.1 & 3.0 & 0.61 & 0.10 & 0.0030 & 0.0017 \\
D1-NVFa & 1 & 1 & 2.68 & $2.3\times10^5$ & 4.0 & 2.7 & 2.1 & 3.0 & 0.88 & 0.10 & \nodata & \nodata \\
 & 1 & 1 & 5.37 & $5.6\times10^4$ & 3.6 & 2.4 & 2.0 & 2.5 & 1.5 & 0.26 & \nodata & \nodata \\
 & 1 & 1 & 10.7 & $1.4\times10^4$ & 3.2 & 2.2 & 1.8 & 1.8 & 2.6 & 0.56 & \nodata & \nodata \\
D1-NVFb & 1 & 1 & 10.7 & $1.4\times10^4$ & 3.2 & 2.2 & 1.8 & 1.8 & 2.7 & 0.60 & \nodata & \nodata \\
 & 1 & 1 & 21.5 & $3.5\times10^3$ & 2.8 & 1.9 & 1.7 & 0.8 & 3.9 & 0.69 & \nodata & \nodata \\
 & 1 & 1 & 43.0 & $880$ & 2.2 & 1.6 & 1.5 & 0 & 4.1 & 0.55 & \nodata & \nodata \\
 & 1 & 1 & 86.0 & 220 & 1.4 & 1.4 & 1.4 & 0 & \nodata & \nodata & \nodata & \nodata
 \enddata
\tablenotetext{a}{Ideal MHD simulations}
\label{tab:param}
\end{deluxetable*}

\subsection{Characterization of the Vertical Structure}\label{sec:vertical}
{ 
We have predicted in Section~\ref{sec:gravity} that 
the relation between the density fluctuation amplitude and 
associated random force depends on the vertical extent of a dead zone.
In order to verify this, we need to define the dead zone in advance and
in a way that does not depend on any specific choice of the model parameters.
In this study, we follow \citetalias{OH11} and define dead and active zones 
in terms of the linear perturbation theory of MRI with Ohmic resistivity.
} 

\begin{figure}
\epsscale{1}
\plotone{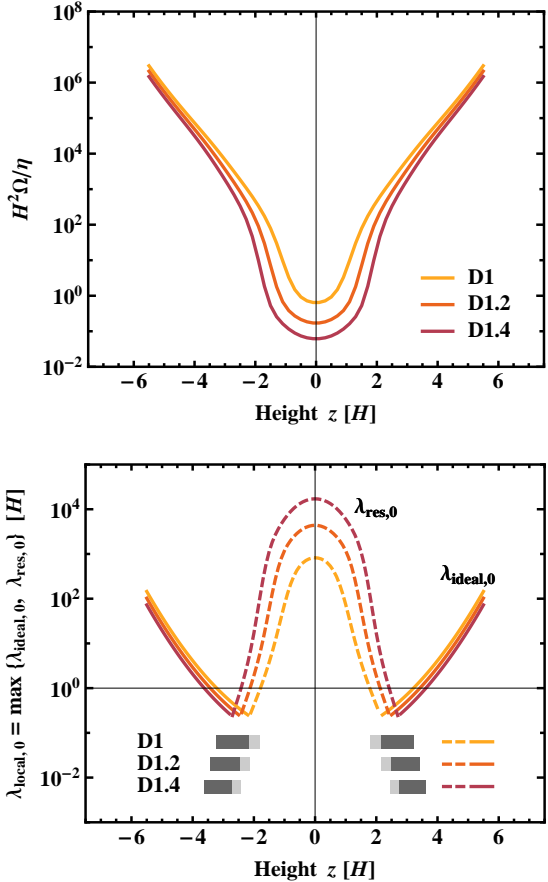}
\caption{Upper panel: magnetic Reynolds number $H^2\Omega/\eta$ vs. height $z$ for models D1, D1.2, and D1.4 of \citetalias{GNT12} (see also their Figure~1).
Lower panel: characteristic MRI wavelength in the laminar state, 
$\lambda_{\rm local,0}$, for the three models.
The solid and dashed segments correspond to $\lambda_{\rm local,0} = \lambda_{\ideal,0}$ (or $\Lambda >1$) and $\lambda_{\rm local,0} = \lambda_{\res,0}$ (or $\Lambda >1$), respectively. 
The vertical line marks $\lambda_{\rm local,0} = H$. 
The dark and light horizontal bars indicate the ideal and resistive MRI regions defined as
$H_{\Lambda,0}<|z|<H_{\ideal,0}$ and $H_{\res,0}<|z|<H_{\Lambda,0}$, respectively (see also Figure~1 of \citetalias{OH11}).
}
\label{fig:z}
\end{figure}

\citetalias{GNT12} calculated the Ohmic resistivity $\eta$
using the charge reaction network {\tt model4} of \citet{IN06a}.
The network consists of free electrons, 
two species of ions (${\rm H_3^+}$ and ${\rm Mg^+}$) and charged dust grains.
We reproduce the resistivity using the analytic prescription presented by \citet{O09} 
with the assumption that ${\rm H_3^+}$ dominates the ions 
at all heights.\footnote{In reality, this assumption is not met at low altitudes 
where the charge transfer from ${\rm H_3^+}$ to ${\rm Mg}$ proceeds rapidly.
However, this hardly affects the ionization degree (and hence the Ohmic resistivity), 
since at the low altitudes the recombination mainly occurs on dust grains, 
for which case the resultant ionization degree is insensitive to the ion composition \citep[see, e.g.,][]{IN06a}.}
To test our calculation, in the upper panel of Figure~\ref{fig:z}, 
we plot the vertical profiles of the magnetic Reynolds number $H^2\Omega/\eta$ 
for models D1, D1.2, and D1.4 of \citetalias{GNT12}.
Here, the disk is assumed to be vertically in hydrostatic equilibrium,  
and the the vertical profile of the disk gas is given by $\rho = \rho_{\rm mid}
\exp(-z^2/2H^2)$, where $\rho_{\rm mid} = \Sigma/\sqrt{2\pi}H$
is the midplane gas density.
Comparing our plot with Figure~1 of \citetalias{GNT12}, 
we confirm that our calculation successfully reproduces their resistivity.

With the information of $\eta$, 
we characterize the vertical turbulent structure of the disk 
using the four-layer description  proposed by \citet{OH11}.
The characterization is based on the linear stability analysis 
of MRI in vertically stratified disks in the presence of Ohmic resistivity \citep{J96,SM99}.
The linear analysis shows that at each height $z$ the gas motion is unstable
if the wavelength $\lambda_{\rm local}(z)$ of the most unstable local MRI mode
is longer than the gas scale height $H$. 
In the presence of Ohmic resistivity, the most unstable wavelength can be approximately given by 
$\lambda_{\rm local} \approx \max\{\lambda_\ideal,\lambda_\res\}$, 
where
\beq
\lambda_\ideal(z) = 2\pi  \frac{v_{Az}(z)}{\Omega}
\eeq
and
\beq
\lambda_\res(z) = 2\pi  \frac{\eta(z)}{v_{Az}(z)}
\eeq
are the characteristic wavelengths of the unstable modes in the ideal and resistive limits, respectively, with $v_{Az} = B_z/\sqrt{4\pi\rho}$ 
being the vertical component of the Alfv\'{e}n velocity.
The most unstable wavelength can be alternatively written as 
$\lambda_{\rm local} \approx \max\{1,\Lambda^{-1}\}\lambda_\ideal$,
where 
\beq
\Lambda \equiv \frac{v_{Az}^2}{\eta\Omega} = \frac{\lambda_\ideal}{\lambda_\res}
\eeq
is the so-called Elsasser number. 
The growth rate $\nu$ of the local MRI modes is approximately given 
by $\nu \approx \min\{1,\Lambda^{-1}\}\Omega$.
The instability is strong ($\nu \sim \Omega$) when $\Lambda \gg 1$, 
weak ($\nu \ll \Omega$) when $\Lambda \ll 1$, 
and absent when $\lambda_{\rm local} > H$.
Thus, the vertical distributions of $\lambda_{\rm ideal}$ and $\lambda_{\rm res}$
predict at which height the MRI is unstable.

When there is a nonzero net vertical magnetic field $\brac{B_z}$, 
it is useful to evaluate $\lambda_{\rm ideal}$ and $\lambda_{\rm res}$
assuming that the disk is in the laminar state, i.e., assuming that  
$\rho = \rho_{\rm mid} \exp(-z^2/2H^2)$ and $B_z = \brac{B_z}$ at all heights.
We denote them by $\lambda_{\ideal,0}$ and $\lambda_{\res,0}$, respectively.
As an example, the lower panel of Figure~\ref{fig:z} plots 
$\lambda_{\rm local,0} = \max\{\lambda_{\ideal,0},\lambda_{\res,0}\}$ 
as a function of $z$ for models D1, D1.2, and D1.4.
Note that $\lambda_{\rm ideal,0}$ and $\lambda_{\rm res,0}$ are 
increasing and decreasing functions of $|z|$, respectively, 
because $\eta$ decreases with $|z|$ while 
$v_{Az,0} = \brac{B_z}/\sqrt{4\pi\rho}$ increases with $|z|$.
For this reason, the region where the MRI is unstable 
(i.e., $\lambda_{\rm local}<H$) is bounded from both below and above.
In this paper, we will refer to such regions as the active layers.
For models D1, D1.2, and D1.4, the active layers are located 
at $1.8H<|z|<3.2H$, $2.1H<|z|<3.4H$ and $2.5H<|z|<3.6H$, respectively.

Based on the stability criterion outlined above, \citetalias{OH11} introduced three critical heights 
$H_\ideal$, $H_\Lambda$, and $H_\res$ defined by
\beq
\lambda_\ideal(z=H_\ideal) = H,
\label{eq:Hideal_def}
\eeq
\beq
\Lambda(z=H_\Lambda) = 1,
\label{eq:HLambda_def}
\eeq
\beq
\lambda_\res(z=H_\res) = H,
\label{eq:Hres_def}
\eeq
respectively. 
The active layer defined by $\lambda_{\rm local} > H$ has a vertical extent 
$H_{\rm res} < |z| < H_{\rm ideal}$.
MRI is stable in the magnetically dominated atmosphere at $|z| < H_{\rm ideal}$,
and in the high-$\eta$ region at $|z| < H_{\rm res}$. 
In this paper, we define a dead zone as the region $|z| < H_{\rm res}$, i.e., 
$H_{\rm DZ} = H_{\rm res}$.
One can subdivide the active layer into two sublayers $H_\Lambda < |z| < H_{\rm ideal}$ and $H_{\rm res} < |z| < H_\Lambda$, at which MRI operates strongly and weakly, respectively. 

As we did for $\lambda_{\rm ideal}$ and $\lambda_{\rm res}$, 
we define the critical heights in the laminar state 
by $H_{\ideal,0}$,  $H_{\Lambda,0}$, and $H_{\res,0}$. 
For $H_{\ideal,0}$, there is an analytic expression (Equation~(14) of \citetalias{OH11})
\beq
H_{\ideal,0} = \left[2\ln\pfrac{\beta_{z0}}{8\pi^2} \right]^{1/2} H,
\label{eq:Hideal0}
\eeq
where $\beta_{z0} \equiv 8\pi\rho_\md/\brac{B_z}^2$ 
is the midplane plasma beta measured with the net field strength $\brac{B_z}$.
Table~\ref{tab:param} list the values of $H_{\ideal,0}$, $H_{\Lambda,0}$, 
and $H_{\res,0}$ for all the \citetalias{GNT12} simulations.
In the lower panel of Figure~\ref{fig:z}, the dark and light horizontal bars 
indicate the ideal and resistive MRI regions, $H_{\Lambda,0} < |z| < H_{\rm ideal,0}$ and $H_{\rm res,0} < |z| < H_{\Lambda,0}$, respectively,  
for models D1, D1.2, and D1.4.

{ In principle, the critical heights in a turbulent state can differ from 
those in the laminar state since $\lambda_\ideal$ and $\lambda_\res$ 
depend on $\rho$ and $B_z$.
One can see this by comparing the critical heights in the initial (laminar) 
and time-averaged (turbulent) states measured in the \citetalias{OH11} simulations 
(see Tables 1 and 2 of \citetalias{OH11}). 
The difference is the largest for $H_\ideal$ due to 
strong fluctuating magnetic fields at the top of active layers, 
and we will discuss this in more detail in Section~\ref{sec:why}. 
In contrast, the difference is much smaller for $H_\Lambda$, 
and is negligible for $H_\res$, 
because magnetic activity is weak at the boarder 
of dead/active regions. 
} 

\subsection{Random Gravitational Force vs Density Fluctuation}
\label{sec:Fvsdrho}
To test Equation~\eqref{eq:F2}, we compare the rms values of 
the random gravitational force and density fluctuations measured by \citetalias{GNT12}. 
Table~\ref{tab:param} lists the rms amplitudes of the density fluctuations at the midplane 
($\brac{\delta\rho^2}_\md^{1/2}$) 
and azimuthal gravitational force acting on test particles ($\brac{F_\phi^2}^{1/2}$) 
measured in the GNT12 simulations.
These values are taken from Table~3 of \citetalias{GNT12}
where they are given in terms of 
the relative fluctuation $\brac{\delta\rho^2}_\md^{1/2}/\rho_\md$
and the rms torque $\brac{\Gamma^2}^{1/2} \equiv a \brac{F_\phi^2}^{1/2}$, respectively.

\begin{figure}
\epsscale{1}
\plotone{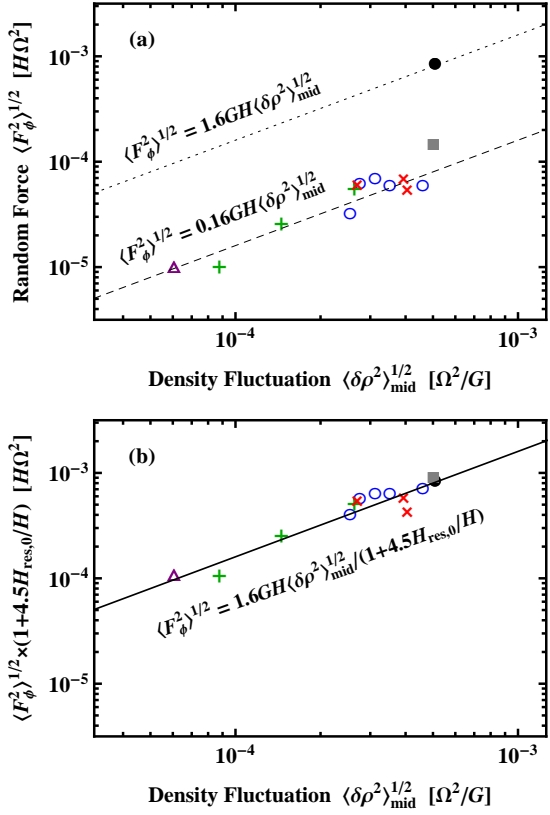}
\caption{Panel (a):  rms azimuthal gravitational force $\brac{F_\phi^2}^{1/2}$ vs.~rms density fluctuation $\brac{\delta\rho^2}^{1/2}_{\rm mid}$ observed in the \citetalias{GNT12} simulations. 
The symbols indicate runs B1 (filled circle), D2 (square), D1 and D1.$x$ (open circles), 
D1-WF (triangle), D1-NVFa (plus signs), and D1-NVFb (crosses). 
The dotted and dashed lines show Equation~\eqref{eq:Fphi1} with ${\cal A}_1 = 1.6$ and $0.16$, respectively. Panel (b): Same as panel (a), but here the force amplitude is rescaled by the factor 
$1+4.5H_{\res,0}/H$ (see also Figure~\ref{fig:A}). 
The solid line indicates Equation~\eqref{eq:Fphi_final}.
}
\label{fig:F}
\end{figure}
Figure~\ref{fig:F}(a) shows $\brac{F_\phi^2}^{1/2}$ versus
$\brac{\delta\rho^2}^{1/2}_\md$ for all the available data.
The dashed and dotted lines in Figure~\ref{fig:F} show 
linear scalings (see Equation~\eqref{eq:F1})
\beq
\brac{F_\phi^2}^{1/2} =  {\cal A}_1GH \brac{\delta\rho^2}_\md^{1/2}
\label{eq:Fphi1}
\eeq 
with ${\cal A}_1 = 1.6$ and 0.16, respectively.
It is clearly seen that no linear scaling can explain the whole data.
\citetalias{GNT12} pointed out this using the data for D1 runs 
(open circles and crosses in Figure~\ref{fig:F}).
We find that this can be seen more clearly by adding 
the data for runs B1 (filled circle) and D2 (filled square), 
for which the dead zone is absent and smaller than 
that in the  D1 runs, respectively.
This fact strengthens the idea that shearing-out of density waves causes
suppression of the random gravity forces. 

\begin{figure}[t]
\epsscale{1}
\plotone{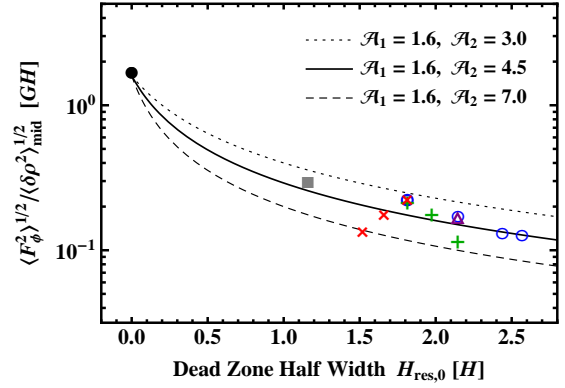}
\caption{Ratio $\brac{F_\phi^2}^{1/2}/\brac{\delta\rho^2}_\md^{1/2}$
vs.~dead-zone half width $H_{\res,0}$. 
The symbols indicate runs B1 (filled circle), D2 (square), D1 and D1.$x$ (open circles), 
D1-WF (triangle), D1-NVFa (plus signs), and D1-NVFb (crosses). 
The dotted, solid, and dashed curves show Equation~\eqref{eq:Fphi2}
with ${\cal A}_1 = 1.6$ and ${\cal A}_2 = 3.0$, $4.5$, and $7.0$, respectively.
}
\label{fig:A}
\end{figure}
From the analysis in Section~\ref{sec:gravity}, we expect that 
the effect of the shearing-out can be extracted by taking the ratio
between $\brac{F_\phi^2}^{1/2}$ and $\brac{\delta\rho^2}_\md^{1/2}$
and comparing it with the half width of the dead zone.
{ As we stated in Section~\ref{sec:vertical}, we measure the dead zone half width 
with the critical height $H_{\rm res}$ defined by Equation~\eqref{eq:Hres_def}.
Specifically, we here use the value in the laminar state, $H_{\rm res,0}$, 
so that we can calculate it directly from the initial conditions
(as we noted in Section~\ref{sec:vertical}, the value of $H_{\rm res}$
is very insensitive to the presence or absence of turbulence). }
Figure~\ref{fig:A} plots the ratio 
$\brac{F_\phi^2}^{1/2}/\brac{\delta\rho^2}_\md^{1/2}$ 
versus $H_{\res,0}$.
We see a decreasing trend in $\brac{F_\phi^2}^{1/2}/\brac{\delta\rho^2}_\md^{1/2}$ 
with increasing $H_{\res,0}$, which is consistent with the idea that 
the random force is weakened as the density waves travel from the active layer 
to the midplane (see Section~\ref{sec:gravity}).
Following Equation~\eqref{eq:F2}, we fit the data shown in Figure~\ref{fig:A}
with a function of the form
\beq
\frac{\brac{F_\phi^2}^{1/2}}{\brac{\delta\rho^2}_\md^{1/2}} 
=  \frac{{\cal A}_1 GH}{1+{\cal A}_2H_{\rm res,0}/H},
\label{eq:Fphi2}
\eeq
where ${\cal A}_1$ and ${\cal A}_2$ are the fitting parameters.
In Figure~\ref{fig:A}, the dotted, solid, and dashed curves show
Equation~\eqref{eq:Fphi2} with ${\cal A}_1 = 1.6$ with 
${\cal A}_2 = 3.0$, 4.5, and 7.0, respectively.
We find that the set $({\cal A}_1, {\cal A}_2) = (1.6, 4.5)$ best reproduces 
the relation between $\brac{F_\phi^2}^{1/2}$ and $\brac{\delta\rho^2}_\md^{1/2}$, within an accuracy of factor 2. 

Thus, we have found that the relation between $\brac{F_\phi^2}^{1/2}$ and $\brac{\delta\rho^2}_\md^{1/2}$ can be well represented by 
\beq
\brac{F_\phi^2}^{1/2} =  \frac{1.6 GH}{1+4.5H_{\rm res,0}/H}
\brac{\delta\rho^2}_\md^{1/2},
\label{eq:Fphi_final}
\eeq
which is also shown in the lower panel of Figure~\ref{fig:F}.

\subsection{Diffusion Coefficients vs Random Force Amplitude}
\label{sec:DvsF}
The next step is to verify the linear scaling between the diffusion coefficients 
and $\brac{F_\phi^2}^{1/2}$ as predicted by Equations~\eqref{eq:Da} 
and \eqref{eq:De}.
\citetalias{GNT12} measured the change in the semimajor axis and eccentricity of
particles without gas friction and with initial eccentricity $e_0 = 0$.
They showed that the ensemble averages of 
$(\Delta a)^2$ and $e^2$ grow linearly with time $\Delta t$,
indicating a random walk of the particles' motion in the phase space. 
\citetalias{GNT12} expressed the time evolution of $(\Delta a)^2$ and $e^2$ 
in the forms
\beq
\brac{(\Delta a)^2}^{1/2}  = C_\sigma(\Delta x) H \pfrac{\Omega \Delta t}{2\pi}^{1/2},
\label{eq:da2dt}
\eeq
\beq
\brac{e^2}^{1/2} = C_\sigma(e)\frac{H}{a} \pfrac{\Omega \Delta t}{2\pi}^{1/2},
\label{eq:de2dt}
\eeq
respectively, where $C_\sigma(\Delta x)$ and $C_\sigma(e)$ are 
dimensionless coefficients that depend on the adopted disk model.
These values are listed in Table 3 of \citetalias{GNT12}.

The diffusion coefficients $D_a$ and $D_e$ can be read off from 
the values of $C_\sigma(\Delta x)$ and $C_\sigma(e)$.
For $D_a$, we have
\beq
D_a = \frac{\left(C_\sigma(\Delta x)\right)^2}{4\pi}{H^2\Omega},
\eeq
which directly follows from Equations~\eqref{eq:Da_def} and \eqref{eq:da2dt}.
For $D_e$, we need to convert $\brac{e^2}^{1/2}$
to the eccentricity displacement $\brac{(\Delta e)^2}^{1/2}$ for general (nonzero) 
initial eccentricity. 
As shown by \citet{YMM09}, this conversion is given by  
$\brac{(\Delta e)^2}^{1/2} =  \sqrt{2/(4-\pi)}\brac{e^2}^{1/2}$.
Hence, from Equations~\eqref{eq:De_def} and \eqref{eq:de2dt}, we get 
\beq
D_e = \frac{2\left(C_\sigma(e)\right)^2}{4\pi(4-\pi)}\frac{ H^2\Omega}{a^2}.
\eeq
The values of $D_a$ and $D_e$ for all the available data are listed 
in Table~\ref{tab:param}.
The listed values are normalized by $H^2\Omega$ and $H^2\Omega/a^2$, respectively, 
which are the natural units for these diffusion coefficients in local simulations. 
\begin{figure}
\epsscale{1}
\plotone{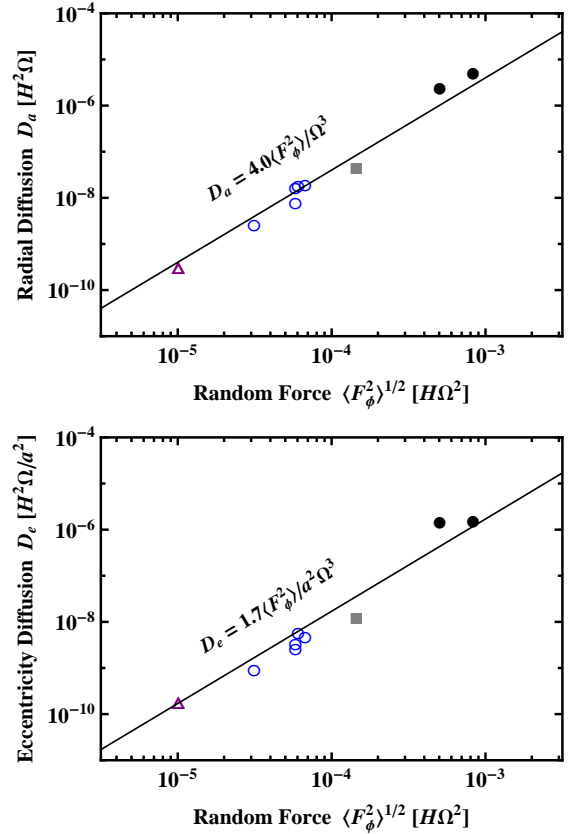}
\caption{Radial diffusion coefficient $D_a$ (upper panel) and eccentricity stirring rate 
$D_e$ (lower panel) vs.~rms azimuthal gravitational force $\brac{F_\phi^2}^{1/2}$ 
for GNT12 simulations. 
The symbols indicate runs A1 and B1 (filled circle), D2 (square), D1 and D1.$x$ (open circles), 
and D1-WF (triangle).
The solid lines in the upper and lower panels show Equation~\eqref{eq:Da} with ${\cal A}_a = 4.0$ and Equation~\eqref{eq:De} with ${\cal A}_e = 1.7$ 
(or equivalently, Equations~\eqref{eq:Da_final} and \eqref{eq:De_final}), respectively.
}
\label{fig:diff}
\end{figure}

Now we calibrate Equations~\eqref{eq:Da} and \eqref{eq:De} 
using the available data.
Figure~\ref{fig:diff} plots $D_a$ and $D_e$ 
versus the rms azimuthal gravitational force $\brac{F_\phi^2}^{1/2}$ 
measured in the \citetalias{GNT12} simulations.
We clearly see the trend $D_a \propto D_e \propto \brac{F^2_\phi}$
as predicted by Equations~\eqref{eq:Da} and \eqref{eq:De}.
We determine the dimensionless parameters ${\cal A}_a$ and ${\cal A}_e$ 
so that the maximum logarithmic error between 
the data and predictions from each of the equations is minimized. 
We find that the best-fit parameters are ${\cal A}_a = 4.0$ and ${\cal A}_e = 1.7$.
The best-fit relations are shown in Figure~\ref{fig:diff}
by the solid lines.

To summarize, we have found that the diffusion coefficients for the semi-major 
axis and eccentricity of solid bodies scale with the mean squared 
amplitude of fluctuating gravity fields as 
\beq
D_a = \frac{4.0 \brac{F_\phi^2}}{\Omega^3},
\label{eq:Da_final}
\eeq
\beq
D_e = \frac{1.7\brac{F_\phi^2}}{a^2\Omega^3}.
\label{eq:De_final}
\eeq
If we eliminate $\brac{F_\phi^2}$ using Equation~\eqref{eq:Fphi2}, 
these scaling relations can be rewritten 
as a function of $\brac{\delta\rho^2}_\md^{1/2}$ and 
$H_{\res,0}$. These read
\beq
D_a 
= \frac{10}{(1+4.5H_{\res,0}/H)^2}\pfrac{a^2H\brac{\delta\rho^2}_\md^{1/2}}{M_*}^2a^2\Omega,
\label{eq:Da_drho}
\eeq
\beq
D_e 
=  \frac{4.4}{(1+4.5H_{\res,0}/H)^2}\pfrac{a^2H\brac{\delta\rho^2}_\md^{1/2}}{M_*}^2\Omega,
\label{eq:De_drho}
\eeq
where we have used that $\Omega = \sqrt{GM_*/a^3}$.

\section{Predicting Density Fluctuation Amplitudes}\label{sec:predictor}
Equations~\eqref{eq:Da_drho} and \eqref{eq:De_drho} tell us how 
the diffusion coefficients $D_a$ and $D_e$ are related to the amplitude 
of the gas density fluctuations, $\brac{\delta\rho^2}_\md^{1/2}$.
To predict the values of $D_a$ and $D_e$ for given disk parameters,
we need to know how $\brac{\delta\rho^2}_\md^{1/2}$ depends on 
these parameters. 

 \citetalias{OH11} provided a simple analytic formula (called the ``saturation predictor'') that predicts $\brac{\delta\rho^2}_\md^{1/2}$ 
as a function of key disk parameters.
In this section, we test whether the formula accurately predicts 
the density fluctuation amplitude observed in the \citetalias{GNT12} simulations.

\subsection{The OH11 Predictor versus GNT12 Data}\label{sec:drho_OH11}
The \citetalias{OH11} saturation predictor reads
\beq
[\brac{\delta\rho^2}^{1/2}_\md]_{\rm OH11}
= \sqrt{0.47\alpha_{\rm core}}\rho_\md,
\label{eq:drho_OH11}
\eeq
where 
$\alpha_{\rm core}$ is a dimensionless coefficient that is
proportional to the turbulent accretion stress integrated over height $|z|< H_\ideal$
(for details, see \citetalias{OH11}).
\citetalias{OH11} provided an empirical formula that relates $\alpha_{\rm core}$ to key disk parameters such as the net vertical flux $\brac{B_z}$ and the vertical distribution of the resistivity $\eta(z)$.
The formula reads
\beq
\alpha_{\rm core} = \frac{510}{\beta_{z0}}
\exp\left(-\frac{0.54H_{\res,0}}{H}\right) 
+ 0.011 \exp\left(-\frac{3.6H_{\Lambda,0}}{H}\right). 
\label{eq:alphacore}
\eeq
In fact, the numerical prefactor appearing in Equation~\eqref{eq:drho_OH11} 
weakly depends on the numerical resolution adopted in simulations;
below we will refine the prefactor in accordance with the data of \citetalias{GNT12} 
simulations that adopted a higher resolution.

\begin{figure}
\epsscale{1}
\plotone{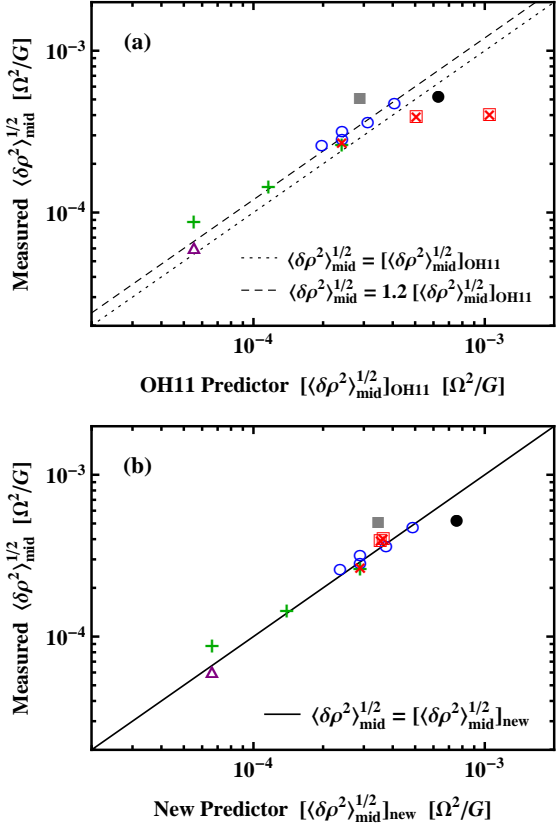}
\caption{Panel (a): midplane density fluctuation $\brac{\delta\rho^2}^{1/2}_\md$ measured in 
\citetalias{GNT12} simulations vs.~the \citetalias{OH11} predictor $[\brac{\delta\rho^2}^{1/2}_\md]_{\rm OH11}$
(Equation~\eqref{eq:drho_OH11}). 
The symbols correspond to runs B1 (filled circle), D2 (filled square), D1 and D1.$x$ (open circles), 
D1-WF (triangle), D1-NVFa (plus signs), and D1-NVFb (crosses). 
The dotted and dashed lines indicate 
$\brac{\delta\rho^2}^{1/2}_\md = [\brac{\delta\rho^2}^{1/2}_\md]_{\rm OH11}$
and $\brac{\delta\rho^2}^{1/2}_\md = 1.2[\brac{\delta\rho^2}^{1/2}_\md]_{\rm OH11}$, respectively.
The two open squares mark the runs 
for which $H_{\ideal,\infty}$ falls below $H_{\res,0}$.
Panel (b): same as panel (a), but here the measured $\brac{\delta\rho^2}^{1/2}_\md$ 
are compared with the updated saturation predictor $[\brac{\delta\rho^2}^{1/2}_\md]_{\rm new}$.
The solid line shows $\brac{\delta\rho^2}^{1/2}_\md = [\brac{\delta\rho^2}^{1/2}_\md]_{\rm new}$.
}
\label{fig:drho}
\end{figure}
In Figure~\ref{fig:drho}(a), we compare the measured values of $\brac{\delta\rho^2}^{1/2}_\md$ for the \citetalias{GNT12} simulations 
with the prediction from Equation~\eqref{eq:drho_OH11}.
We see that the \citetalias{OH11} predictor reasonably reproduces 
most of the observed values,
especially for runs D1, D1.$x$, D1-WF, and D1-NVFa. 
However, detailed inspection shows that these observed values are 
higher than the prediction by $\approx 20\%$ (see the dotted line in Figure~\ref{fig:drho}(a)).
This discrepancy can be attributed to the fact that \citetalias{GNT12} adopted 
a higher numerical resolution than \citetalias{OH11}.
As shown by both \citetalias{OH11} and \citetalias{GNT12}, 
the density fluctuation amplitude increases slowly with improving the numerical resolution (see Figure 18 of \citetalias{OH11} and Figure A1 of \citetalias{GNT12}).
We expect that the values measured by \citetalias{GNT12} are well converged 
to the true values since an even higher resolution does not give any significant 
change in the density fluctuation amplitude (see Appendix A of \citetalias{GNT12}).

A more important discrepancy can be found for runs D1-NVF.
In these runs, the net vertical field strength $\brac{B_z}$ 
was gradually increased from 2.7 mG to 86.0 mG.
The results show that $\brac{\delta\rho^2}^{1/2}_\md$ initially 
increases with $\brac{B_z}$ but gets suppressed at 
$\brac{B_z} > 10.7~{\rm mG}$. 
This can be seen in Figure 7 of \citetalias{GNT12}, 
and we also show this in the lower panel of our Figure~\ref{fig:B}.
The \citetalias{OH11} predictor does not reproduce this suppression
(as shown by the gray dashed line in Figure~\ref{fig:B}) 
and consequently overestimates the density fluctuation amplitude for 
$\brac{B_z} = 21.5$ and 43.0 mG (marked by the open squares in Figure~\ref{fig:drho}).

\subsection{Why are the Density Fluctuations Suppressed at High $\brac{B_z}$?}
\label{sec:why}
\citetalias{GNT12} explained the suppressed 
$\brac{\delta\rho^2}^{1/2}_\md$ at high $\brac{B_z}$ 
as a consequence of narrowed MRI-active layers.
In a stratified disk, an MRI-active layer is bounded from above 
by a magnetically dominated atmosphere.
As the magnetic fields become stronger, the base of the atmosphere 
moves down to the midplane, narrowing the active layer beneath.
In the limit of high fields, the atmosphere will erode  
most of the active layer, and will in turn suppress 
the generation of density fluctuations.
\begin{figure}
\epsscale{1}
\plotone{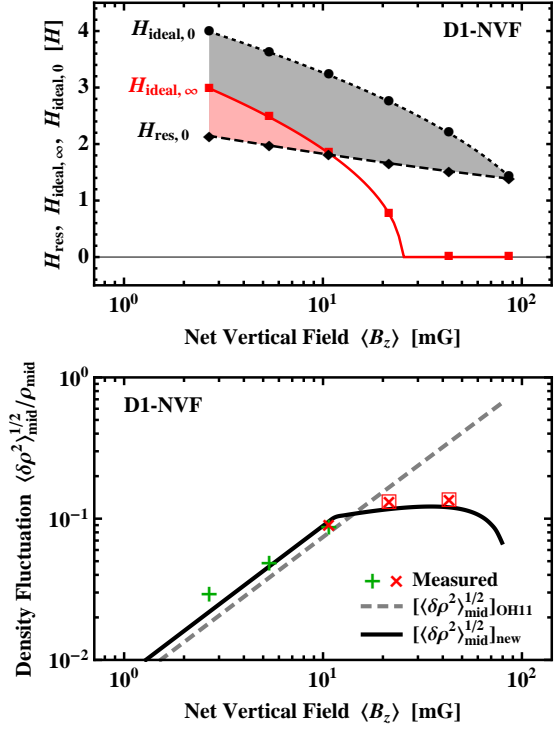}
\caption{Upper panel: critical heights $H_{\ideal,0}$ (dashed curve), $H_{\res,0}$ (dashed curve), 
and $H_{\ideal,\infty}$ (solid curve) as a function of the net vertical field strength $\brac{B_z}$ 
for model D1-NVF.
Lower panel: midplane density fluctuation $\delta\rho_\md/\rho_\md$ measured 
in the D1-NVF runs (plus and cross symbols), compared with 
the \citetalias{OH11} predictor $[\brac{\delta\rho^2}^{1/2}_\md]_{\rm OH11}$ (Equation~\eqref{eq:drho_OH11}; dashed curve)
and the updated saturation predictor $[\brac{\delta\rho^2}^{1/2}_\md]_{\rm new}$ 
(Equation~\eqref{eq:drho_new}; solid curve).
The two open squares mark the runs for which $H_{\ideal,\infty} < H_{\res,0}$.
}
\label{fig:B}
\end{figure}

To confirm the hypothesis raised by \citetalias{GNT12}, we estimate 
the vertical extent of the active layer using the critical heights we introduced in Section~\ref{sec:vertical}.
As explained there, local MRI modes can exist only at $z < H_\ideal$, 
above which $\lambda_{\rm local} (\approx \lambda_\ideal)$ exceeds the gas scale height $H$. 
The region where local MRI modes exist 
is also bounded from below by $z=H_\res$, below which the Ohmic resistivity 
stabilizes all MRI modes (i.e., $\lambda_{\rm local} \approx \lambda_\res > H$).
Thus, we may measure the vertical width of an MRI-active layer as 
$\Delta z_{\rm active} = H_\ideal - H_\res$, with $H_\ideal$ corresponding 
to the base of the magnetically dominated atmosphere
while $H_\res$ to the the active/dead zone interface.

The active layer width defined above naturally explains 
the suppression of the density fluctuation amplitude at high net fields.
In the upper panel of Figure~\ref{fig:B}, we plot the critical heights 
in the laminar state, $H_{\ideal,0}$ and $H_{\res,0}$, 
as a function of $\brac{B_z}$.
We see that the active layer width in the laminar state 
$\Delta z_{\rm active,0} = H_{\ideal,0}-H_{\res,0}$ is already 
as small as $\la H$ for $\brac{B_z} \ga 10~{\rm mG}$.
Furthermore, $\Delta z_{\rm active,0}$ vanishes 
at $\brac{B_z} \approx 80~{\rm mG}$, indicating that 
the magnetically dominated atmosphere completely suppresses 
the MRI-active layer for this value of $\brac{B_z}$ or larger.
This exactly explains what happened in run D1-NVFb, 
in which the disk returned to a laminar state 
at  $\brac{B_z} = 86~{\rm mG}$ (see Figure~11 of \citetalias{GNT12}).

A more quantitative analysis can be made by noting 
that $H_\ideal$ further decreases as MRI-driven turbulence develops.
Simulations by \citetalias{OH11} show that $H_\ideal$ measured in the fully turbulent state 
is smaller { than} that in the laminar state ($H_{\ideal,0}$) by about one scale height (see their Tables 1 and 2).
By contrast, $H_\res$ is nearly independent of the turbulence state in the active layer because 
turbulence is always weak on the dead/active zone boundary.
The definition of $H_\ideal$ (Equation~\eqref{eq:Hideal_def}) allows us to directly
calculate how much $H_\ideal$ decreases as the turbulence develops 
at the top of the active layer.
If we measure $v_{Az}$ with the rms amplitude of the $B_z$ fields, $\brac{B_z^2}^{1/2}$, then Equation~\eqref{eq:Hideal_def} can be rewritten as 
\beq
\rho_\ideal c_s^2 = \pi \brac{B_z^2}_\ideal
\label{eq:ideal}
 \eeq
where $\brac{B_z^2}_\ideal$ and $\rho_\ideal$ are the values of 
$\brac{B_z^2}$ and $\rho$ at $z = H_\ideal$, respectively.
We may approximate $\rho_\ideal$ with the hydrostatic density profile, 
$\rho_\ideal \approx \rho_\md\exp(-H_\ideal^2/2H^2)$, since the gas pressure dominates 
over the magnetic pressure at $z\leq H_\ideal$ (see also Figure~4(a) of \citetalias{OH11}).
Thus, the definition of $H_\ideal$  can be further rewritten as
\beqn
H_\ideal^2 &=& 2H^2 \ln\pfrac{\rho_\md c_s^2}{\pi \brac{B_z^2}_\ideal} 
\nonumber \\
&=& H_{\ideal,0}^2 - 2H^2\ln\left(1 + \frac{\brac{\delta B_{z}^2}_\ideal}{\brac{B_z}^2} \right),
\label{eq:Hideal}
\eeqn
where $H_{\ideal,0}$ is the value of $H_\ideal$ in the laminar state (Equation~\eqref{eq:Hideal0})
and $\brac{\delta B_{z}^2}_\ideal \equiv \brac{B_z^2}_\ideal  - \brac{B_z}^2 $
 is the mean squared amplitude of the fluctuating (turbulent) $B_z$-fields at $z = H_\ideal$.
Equation~\eqref{eq:Hideal} demonstrates that $H_\ideal$ decreases as turbulence grows.

It is useful to know to what extent $H_{\ideal}$ decreases 
{\it if turbulence is fully developed}.
The results of the \citetalias{OH11} simulations show that 
in the fully turbulent state, $\brac{\delta B_z^2}_\ideal$ 
approximately satisfies the relation
(see Appendix~\ref{sec:OH11})
\beq
\brac{\delta B_{z}^2}_\ideal \sim 30 \brac{B_z}^2.
\label{eq:dBz2_ideal}
\eeq
Inserting this relation into Equation~\eqref{eq:Hideal}, we find that 
$H_\ideal$ should decrease to $H_{\ideal,\infty}$, where  
\beqn
H_{\ideal,\infty}^2 &\approx& H_{\ideal,0}^2 - 7H^2.
\label{eq:Hideal_infty}
\eeqn

In the upper panel of Figure~\ref{fig:B}, the solid curve shows 
$H_{\ideal,\infty}$ versus $\brac{B_z}$ for model D1-NVF. 
The values of $H_{\ideal,\infty}$ for all the GNT12 simulations are listed in Table 1.
Since $H_\res$ is hardly affected by turbulence, we can estimate the width of the active layer in the fully turbulent state as 
$\Delta z_{\rm active,\infty} = H_{\ideal,\infty} - H_{\res,0}$.
As seen in the figure, $\Delta z_{\rm active,\infty}$ becomes negative for $\brac{B_z} \ga 10~{\rm mG}$.
This suggests that {\it turbulence cannot be fully developed} in run D1-NVFb with 
$\brac{B_z} = 21.5$ and 43.0 mG; if turbulence were fully developed, then the magnetically dominated atmosphere would completely suppress the active layer.

\begin{figure*}
\epsscale{1.15}
\plotone{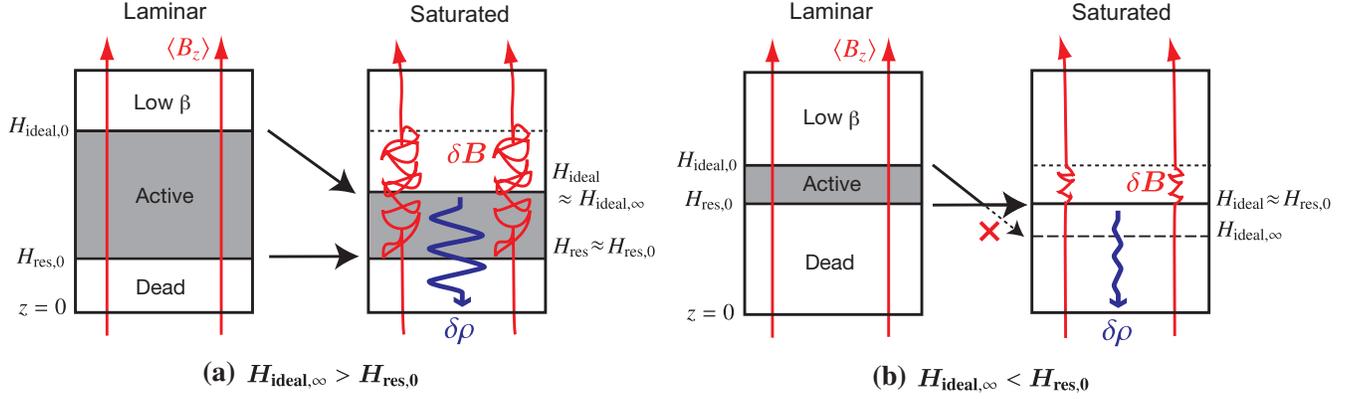}
\caption{Schematic description of the physics behind the saturation limiter
${\cal L}$ (Equation~\eqref{eq:L}).
The gray regions indicate the MRI-active layer 
defined by $H_{\rm res}<z<H_{\rm ideal}$.
The upward arrows show the magnetic fields, while the downward 
wavy arrows represent propagation of density waves from the active layer 
to the midplane.
In the laminar state, the upper boundary of the active layer is located at 
$z=H_{\rm ideal,0}$ (Equation~\eqref{eq:Hideal0}).
As turbulence develops, $H_{\rm ideal}$ decreases because of the buildup 
of fluctuating $B_z$ fields, and reaches $H_{\rm ideal,\infty}$
(Equation~\eqref{eq:Hideal_infty}) when the turbulence is fully developed. 
If $H_{\rm ideal,\infty} > H_{\rm res}$ (case (a)), the active layer can have a 
finite thickness even with $H_{\rm ideal} = H_{\rm ideal,\infty}$, and hence 
a fully turbulent state is realized (${\cal L} = 1$).
If $H_{\rm ideal,\infty} < H_{\rm res}$ (case (b)), the turbulence stops developing  
at the point where $H_{\rm ideal}$ reaches $H_{\rm res}$, and hence 
gets saturated at a low level (${\cal L} < 1$).
}
\label{fig:model}
\end{figure*}

The above analysis confirms the hypothesis by \citetalias{GNT12} 
that the magnetically dominated atmosphere 
limits the saturation level of turbulence in the active layer.
We summarize the mechanisms of this effect in Figure~\ref{fig:model}.
This effect is not taken into account in the \citetalias{OH11} saturation predictor (Equation~\eqref{eq:drho_OH11}) as $H_{\rm ideal,\infty}$ was larger than 
$H_{\rm res,0}$ for all the \citetalias{OH11} simulations.
This explains why the \citetalias{OH11} predictor overestimates 
the amplitude of the density fluctuations at high $\brac{B_z}$.

\subsection{Refining the OH11 Predictor with the ``Saturation Limiter''}
\label{sec:drho_new}
Based on the above consideration, we construct 
a toy model that accounts for the suppression 
of the density fluctuation amplitude at high $\brac{B_z}$.

The model is based on two assumptions.
Firstly, we assume that 
if $H_{\ideal,\infty} < H_{\res,0}$ then
turbulence grows until $H_{\ideal}$ reaches $H_{\res,0}$. 
This means that the saturated value of $\brac{\delta B_{z}^2}_\ideal$
is given by Equation~\eqref{eq:Hideal} with $H_\ideal = H_{\res,0}$.
Solving the equation with respect to $\brac{\delta B_{z}^2}_\ideal$, we get 
\beq
\brac{\delta B_{z}^2}_\ideal = \left[\exp\pfrac{H_{\ideal,0}^2 - H_{\res,0}^2}{2H^2}-1\right]
\brac{B_z}^2.
\label{eq:dBz_saturated}
\eeq
Combining this with Equation~\eqref{eq:dBz2_ideal}, 
we obtain the saturation predictor for $\brac{\delta B_{z}^2}_\ideal$
for general cases,
\beq
\brac{\delta B_{z}^2}_\ideal = 30{\cal L}\brac{B_z}^2,
\label{eq:dBz_limit}
\eeq
where ${\cal L}$ is defined by
\beq
{\cal L}= \min\biggl\{1, \frac{1}{30}\left[\exp\left(\frac{H_{\ideal,0}^2 - H_{\res,0}^2}{2H^2}\right)-1\right]\biggr\}.
\label{eq:L}
\eeq
This expresses that turbulence in the active layer is limited at a low level 
(${\cal L} < 1$) when $H_{\ideal,\infty} < H_{\res,0}$ (see also Figure~\ref{fig:model}).
We will call ${\cal L}$ the ``saturation limiter.''

Secondly, we assume that 
$\brac{\delta\rho^2}_\md$ is proportional to $\brac{\delta B_z^2}_\ideal$;
namely, if $\brac{\delta B_{z}^2}_\ideal$ is suppressed by factor ${\cal L}$, 
$\brac{\delta\rho^2}_\md$ is suppressed by the same factor. 
This assumption can be expressed as
\beq
\brac{\delta\rho^2}_\md = {\cal L}\brac{\delta\rho^2}_{\md,\infty},
\label{eq:drho_limit}
\eeq
where $\brac{\delta\rho^2}_{\md,\infty}$ is the value of 
$\brac{\delta\rho^2}_{\md}$ for fully developed MRI turbulence 
(${\cal L} = 1$).
We take $\brac{\delta\rho^2}_{\md,\infty}^{1/2} = 1.2[\brac{\delta\rho^2}^{1/2}_\md]_{\rm OH11}$ based on the results of the \citetalias{GNT12} simulations
for low $\brac{B_z}$ (see Section~\ref{sec:drho_OH11}).
This second assumption can be validated with the following argument. 
\citetalias{OH11} showed that that the internal energy density of fluctuation, 
$c_s^2\brac{\delta\rho^2}/2\rho$, is nearly constant 
along the vertical direction (see also Figure~13 of \citetalias{GNT12}). 
This implies that
\beq
\frac{\brac{\delta\rho^2}_\md}{\rho_\md} \sim 
\frac{\brac{\delta\rho^2}_\ideal}{\rho_\ideal}.
\label{eq:Eint}
\eeq
It can also be shown that the internal and magnetic energy densities 
of fluctuation are nearly equipartitioned at $|z|\approx H_\ideal$, i.e., 
\beq
\frac{c_s^2\brac{\delta\rho^2}_\ideal}{2\rho_\ideal} \sim \frac{\brac{\delta B_{z}^2}_\ideal}{8\pi}
\label{eq:equip}
\eeq 
(see Appendix~\ref{sec:OH11} for the supporting data).
Equations~\eqref{eq:Eint} and \eqref{eq:equip} imply that 
$c_s^2\brac{\delta\rho^2}_\md/2\rho_\md \sim \brac{\delta B_z^2}_\ideal/8\pi$,
and hence $\brac{\delta\rho^2}_\md \propto \brac{\delta B_z^2}_\ideal$.

Substituting $\brac{\delta\rho^2}_{\md,\infty}^{1/2} = 1.2[\brac{\delta\rho^2}^{1/2}_\md]_{\rm OH11}$ and Equation~\eqref{eq:drho_OH11}  
 into Equation~\eqref{eq:drho_limit}, 
we arrive at the new saturation predictor
\beq
[\brac{\delta\rho^2}^{1/2}_\md]_{\rm new} =  \sqrt{0.68{\cal L}\alpha_{\rm core}}\rho_\md,
\label{eq:drho_new}
\eeq
where the factor $\sqrt{0.68}$ comes from $1.2\times\sqrt{0.47}$.
{ We stress again that the saturation limiter is relevant only when 
the net flux is so high as to significantly change $H_{\ideal}$.
If the net field is not so strong and hence ${\cal L}=1$ holds, 
then the new predictor differs from 
the previous \citetalias{OH11} predictor only by factor $1.2$.
}

To see how the saturation limiter ${\cal L}$ operates at high $\brac{B_z}$, 
we compare in the bottom panel of Figure~\ref{fig:B}
the midplane density fluctuation 
measured in runs D1-NVF with the \citetalias{OH11} predictor $[\brac{\delta\rho^2}^{1/2}_\md]_{\rm OH11}$
and updated predictor $[\brac{\delta\rho^2}^{1/2}_\md]_{\rm new}$.
Comparison with the whose data are shown in Figure~\ref{fig:drho}(b).
As we see, $[\brac{\delta\rho^2}^{1/2}_\md]_{\rm OH11}$ predicts a monotonic increase 
in the fluctuation amplitude toward high $\brac{B_z}$, 
while $[\brac{\delta\rho^2}^{1/2}_\md]_{\rm new}$ predicts a flat amplitude at $\brac{B_z} \ga 10~{\rm mG}$
because of the saturation limiter.
The prediction by $[\brac{\delta\rho^2}^{1/2}_\md]_{\rm new}$ 
is remarkably consistent with the measurements by \citetalias{GNT12}.
{ The new predictor also naturally explains the fact that 
the disk returns to a laminar state (i.e., the density fluctuations vanish) 
at $\brac{B_z} = 86~{\rm mG}$.
We stress that the saturation limiter has been constructed 
{\it without further calibration with the numerical data of \citetalias{GNT12}}.
It is remarkable that it nevertheless explains the observed tendency of 
the density fluctuation amplitude at high net vertical fields. 
However, given the limited number of the data points, 
further support from simulations is desirable 
to warrant its general validity.
}

The same refinement should be applicable to 
the saturation predictor for the gas velocity fluctuation amplitude 
$\brac{\delta v^2}^{1/2}$ since $\delta v \propto \delta \rho$ for sound waves.
The \citetalias{OH11} predictor for the midplane velocity fluctuation amplitude 
is given by
\beq
[\brac{\delta v^2}^{1/2}_{\md}]_{\rm OH11} =  \sqrt{0.78 \alpha_{\rm core}}c_s.
\eeq
As we did for the density fluctuation amplitude, 
we multiply $[\brac{\delta v^2}^{1/2}_{\md}]_{\rm OH11}$ by $1.2\sqrt{{\cal L}}$
to obtain a new predictor 
\beq
[\brac{\delta v^2}^{1/2}_{\md}]_{\rm new} =  \sqrt{1.1{\cal L}\alpha_{\rm core}}c_s.
\eeq

It should be noted, however, that the accretion stress (or effective viscosity) 
is not necessarily limited since non-fluctuating, large-scale magnetic 
fields can contribute to it \citep{TS08,GNT11,BS13a,BS13b}.
Indeed, \citetalias{GNT12} observed no suppression in the accretion stress 
at high $\brac{B_z}$ where the density fluctuation amplitude already 
reaches the ceiling (see their Figure~9).

\section{Discussion}\label{sec:discussion}

\subsection{The Predictor Functions for the Diffusion Coefficients}
In Section~\ref{sec:calibration}, we have found how 
the diffusion coefficients for the orbital elements of planetesimals, 
$D_a$ and $D_e$, are related to the amplitude of
density fluctuations at the midplane, $\brac{\delta \rho^2}_\md^{1/2}$. 
We have also obtained in Section~\ref{sec:predictor} the saturation predictor for $\brac{\delta \rho^2}_\md^{1/2}$
as a function of the net vertical flux $\brac{B_z}$ 
and vertical distribution of the Ohmic resistivity $\eta(z)$.
Here, we combine these relations to 
provide the predictor functions for $D_a$ and $D_e$. 
Substitution of Equation~\eqref{eq:drho_new} into Equations~\eqref{eq:Da_drho} 
and \eqref{eq:De_drho} gives
\beq
D_a = \frac{1.1 {\cal L}\alpha_{\rm core}}{(1+4.5H_{\res,0}/H)^2}\pfrac{\Sigma a^2}{M_*}^2 a^2\Omega,
\label{eq:Da_pred}
\eeq
\beq
D_e = \frac{0.47 {\cal L}\alpha_{\rm core}}{(1+4.5H_{\res,0}/H)^2}\pfrac{\Sigma a^2}{M_*}^2 \Omega,
\label{eq:De_pred}
\eeq
respectively, where we have used that $\rho_\md = \Sigma/\sqrt{2\pi}H$.
The factor ${\cal L}\alpha_{\rm core}$ come from the fact  
that the diffusion coefficients are proportional to $\brac{\delta \rho^2}_\md$.
The factor $(1+4.5H_{\res,0}/H)^{-2}$ accounts for 
the suppression of the random gravity due to the shearing-out 
of the density fluctuations in the presence of a dead zone. 
These formulae together with the predictor for $\alpha_{\rm core}$ (Equation~\eqref{eq:alphacore}) allow us to compute the turbulent diffusion coefficients 
as a function of the net vertical flux and vertical distribution of Ohmic resistivity.

\subsection{Comparison with Previous Recipes Based on Ideal MHD Simulations}\label{sec:comparison}
We check the consistency between our stirring recipe and previous ones proposed 
by \citet{IGM08} and \citet{YMM09,YMM12}.
The previous recipes assume ideal MHD, 
so we will take $H_{\res,0} = H_{\Lambda,0} = 0$ in the following comparison.
Then, Equations~\eqref{eq:Da_pred} and \eqref{eq:De_pred} reduce to
\beq
D_a \approx 5.5\times 10^{-3}\pfrac{\alpha}{10^{-2}} \pfrac{\Sigma a^2}{M_*}^2 a^2\Omega,
\label{eq:Da_ideal}
\eeq
\beq
D_e \approx 2.4\times 10^{-3}\pfrac{\alpha}{10^{-2}} \pfrac{\Sigma a^2}{M_*}^2 \Omega,
\label{eq:De_ideal}
\eeq
respectively.
Here, we have used that $\alpha_{\rm core} \approx \alpha/2$ in the absence of a dead zone, where $\alpha$ is the Shakura--Sunyaev viscosity parameter.\footnote{{ By definition, $\alpha$ is the sum of $\alpha_{\rm core}$ 
and $\alpha_{\rm atm}$, where the latter is related to the accretion stress in the magnetized atmosphere ($|z| > H_\ideal$) and has little effect on density/velocity fluctuations near the midplane. For details, see Section~5 of \citetalias{OH11}.}}
The above forms are useful when comparing the above equations with previous recipes.

\subsubsection{Ida et al. (2008)}\label{sec:IGM08}
\citet{OIM07} and \citet{IGM08} simulated 
turbulent stirring of planetesimals using a random gravity field model
that mimics MRI-driven turbulence originally proposed by \citet{LSA04}.
Based on the results of these simulations, \citet{IGM08} 
proposed a simple formula for the eccentricity growth, 
\beq
\brac{e^2}^{1/2} \approx 0.1\gamma \pfrac{M_\odot}{M_*}\pfrac{\Sigma}{2400~{\rm g~cm^{-2}}}
\pfrac{a}{1~\AU}^2 \pfrac{\Omega \Delta t}{2\pi}^{1/2},
\label{eq:e_IGM08}
\eeq
where $\gamma$ is a dimensionless parameter that characterizes 
the amplitude of the modeled random gravity fields \citep[see Equation~(6) of][]{OIM07}.
Here, we augmented the factor $M_\odot/M_*$ to the original formula,
since \citet{IGM08} fixed $M_*$ to be $M_\odot$
while the magnitude of the modeled random gravity 
actually scales as $M_*^{-1}$ \citep[see Equation~(5) of][]{OIM07}. 
\citet{IGM08} expected $\gamma \sim 10^{-2}$--$10^{-3}$ 
for ideal MRI-turbulent disks from the results 
of MHD simulations by \citet{LSA04}.

To enable comparison,  
we rewrite Equation~\eqref{eq:e_IGM08} in terms 
of the eccentricity diffusion rate defined by Equation~\eqref{eq:De_def}.
If we use Equation~\eqref{eq:De} together with the relationship 
$\brac{e^2} \sim \brac{(\Delta e)^2}$ (see Section~\ref{sec:DvsF}), 
the above formula can be rewritten as  
\beq
[D_e]_{\rm IGM08} \approx 0.01 \pfrac{\gamma}{10^{-3}}^2 
\pfrac{\Sigma a^2}{M_*}^2 \Omega.
\label{eq:De_IGM08}
\eeq
Comparison between $[D_e]_{\rm IGM08}$ and our $D_e$ 
(Equation~\eqref{eq:De_ideal}) allows us to know how 
the dimensionless parameter $\gamma$ should be related 
to the strength of turbulence, $\alpha$. 
We find  
\beq
\gamma \approx 5\times 10^{-4}\pfrac{\alpha}{10^{-2}}^{1/2}.
\label{eq:gamma}
\eeq  
As we will see in Section~\ref{sec:YMM12}, 
\citet{YMM12} obtained a consistent result for the case of 
$\alpha \sim 10^{-2}$. \citet{BL10} also obtained a similar conclusion
based on hydrodynamical simulations with \citet{LSA04}'s random gravity model.
In the absence of a dead zone, $\alpha$ takes a value of $\sim 10^{-2}$ 
or larger depending on the strength of the net vertical magnetic flux 
\citep{DSP10,SMI10}. 
Therefore, our recipe suggests that $\gamma \ga 10^{-3}$ 
for ideal MRI-turbulent disks,
supporting the expectation by \citet{IGM08}. 

However, we stress again that a simple relationship between $\gamma$ 
and $\alpha$ like Equation~\eqref{eq:gamma} does not  apply 
in the presence of a dead zone.
Comparison between Equations~\eqref{eq:De_pred} and \eqref{eq:De_IGM08}
shows that $\gamma$ must be interpreted as 
\beq
\gamma \approx \frac{7\times 10^{-4}{\cal L}^{1/2}}{1+4.5H_{\rm res,0}/H}
\pfrac{\alpha_{\rm core}}{10^{-2}}^{1/2}.
\label{eq:gamma_eff}
\eeq
The factor $\alpha_{\rm core}$ express the level of the accretion stress
at low altitudes and therefore crudely corresponds to $\alpha$ in Equation~\eqref{eq:gamma}.
A dead zone reduces the value of $\alpha_{\rm core}$ as expressed by 
Equation~\eqref{eq:alphacore}. 
However, this factor does not capture all the roles of a dead zone.
A dead zone induces the shearing-out of density waves, 
and thereby further reduces the planetesimal stirring rate 
as expressed by the prefactor $(1+4.5H_{\rm res,0}/H)^{-1}$ 
in Equation~\eqref{eq:gamma_eff}.
Suppression of MRI activity at very high $\brac{B_z}$ 
also reduces $\gamma$ through the saturation limiter ${\cal L}$.

\subsubsection{Yang et al. (2012)}\label{sec:YMM12}
\citet{YMM09,YMM12} studied planetesimal stirring in local ideal MHD simulations,
with an emphasis on the dependence of the results 
on the horizontal box size adopted in the simulations.
For stratified disks, \citet{YMM12} proposed analytic expressions 
for $\brac{(\Delta a)^2}^{1/2}$ and $\brac{(\Delta e)^2}^{1/2}$,
\beq
\brac{(\Delta a)^2}^{1/2} = 6.6\times 10^{-5}\pfrac{L_h}{\sqrt{2}H}^{1.35} \xi  \sqrt{2}H\pfrac{\Omega \Delta t}{2\pi}^{1/2},
\eeq
\beq
\brac{(\Delta e)^2}^{1/2} = 7.2\times 10^{-5}\pfrac{L_h}{\sqrt{2}H}^{1.08} \xi \frac{\sqrt{2}H}{a}\pfrac{\Omega \Delta t}{2\pi}^{1/2},
\eeq
where $\xi \equiv 4\pi G \rho_\md ({2\pi}/{\Omega})^2$ and 
$L_h$ is the horizontal box size.
In terms of the diffusion coefficients (Equation~\eqref{eq:Da_def} and \eqref{eq:De_def}), 
these expressions can be written as 
\beq
[D_a]_{\rm YMM12} \approx 0.03 \pfrac{L_h}{20H}^{2.16}
\pfrac{\Sigma a^2}{M_*}^2 a^2\Omega,
\label{eq:Da_YMM12}
\eeq
\beq
[D_e]_{\rm YMM12} \approx 0.01 \pfrac{L_h}{20H}^{2.7}
\pfrac{\Sigma a^2}{M_*}^2\Omega.
\label{eq:De_YMM12}
\eeq
Comparing the above expressions 
with Equations~\eqref{eq:Da_ideal} and \eqref{eq:De_ideal},
we find that our stirring recipe is reasonably consistent with those of \citet{YMM12}
given that \citetalias{GNT12} adopted the local box of azimuthal extent $12$--$16H$
and that $\alpha \sim 10^{-2}$ in the simulations of \citet[][see their Figure 2, bottom panel]{YMM12}.

We here note that the horizontal box size dependence appearing in 
Equations~\eqref{eq:Da_YMM12} and \eqref{eq:De_YMM12} does not affect 
the validity of our recipe at a practical level.
\citet{YMM12} suggest that 
the box size of local simulations needs to be as large as the orbital radius  
to reproduce the results of global simulations. 
However, the azimuthal box size of the \citetalias{GNT12} simulations was 
as large as the orbital radius, since $a \sim 20H$ in a typical protoplanetary disk.
In fact, \citet{YMM12} suggested $\gamma \approx 6\times 10^{-4}$ 
for $\alpha \sim 10^{-2}$ under the criterion $L_h \approx a$ 
(see their Section~6.2), which is precisely consistent with our conclusion
(Equation~\eqref{eq:gamma}).

Invoking the results of \citet{IGM08} for $\gamma \sim 10^{-3}$, 
\citet{YMM12} concluded that planetesimals are able to survive mutual 
collisional destruction even in fully developed MRI turbulence. 
However, this conclusion should be interpreted with care since 
this is only true in inner regions of protoplanetary disks 
where $a \sim 1~\AU$ \citep[see Figure~4 of][]{IGM08}.
It is also important to note that a high net vertical magnetic flux 
can give a higher $\alpha$ and hence a higher $\gamma$.
Moreover, growth of planetesimals to protoplanets is not obvious 
even in the inner disk regions, because the condition 
for gravitational runaway growth is generally severer than 
that for surviving collisional destruction. 
In fact, according to the results of \citet{IGM08}, 
the growth condition can break down 
in the entire part of protoplanetary disks for $\gamma \sim 10^{-3}$
unless the planetesimal size is larger than 100~km.
We will discuss this issue in more detail in \citetalias{OO13b}.

\subsection{Implications for the Excitation of Planetesimals 
in Protoplanetary Disks}
In disks harboring MRI-inactive dead zones, 
the planetesimal stirring rate $D_e$ 
will be reduced considerably with respect to the ideal MHD limit. 
Two mechanisms contribute to reduce $D_e$:

\begin{enumerate}
\item
The effective $\alpha$, or rather, the amplitude of the density
fluctuations, is smaller when the MRI is not ideal.
\item
The shearing-out will distort the geometry of the density fluctuations,
rendering them less effective to stir planetesimals at the midplane.
\end{enumerate}

The first effect is primarily a function of the net vertical magnetic flux $\brac{B_z}$ 
(see OH11). 
At low $\brac{B_z}$, the magnitude of the density fluctuations becomes low; that is, the disks become less turbulent. However, the density fluctuation amplitude cannot be
too large even for very high $\brac{B_z}$ as the magnetically dominated atmosphere will then suppress the active layer (this is expressed by the saturation limiter ${\cal L}$ we introduced in Section~\ref{sec:drho_new}). 
The second effect is also very important. This will reduce the random torque 
by up to factor $\sim 10$ (see Figure~\ref{fig:A}) 
and consequently the stirring rate by up to $\sim 100$ 
when disk possess dead zones.

In this work we have presented a general machinery to capture these
effects and to quantify the excitation rate $D_e$. 
Unfortunately, in contrast to Equation~\eqref{eq:De_ideal}, 
in the presence of a dead zone the effective excitation rate 
is no longer a simple function of the viscosity $\alpha$. 
Apart from $\brac{B_z}$, 
it also becomes a function of the resistivity profile, $\eta(z)$, 
which reflects the ionization fraction of the gas. 
And to calculate the ionization fraction one requires to know the ionization sources, 
their rates, and the properties of small dust grains.

Previous studies investigating planetesimal accretion 
\citep[e.g.,][]{MBNL09,ODS10b,W11,M12} have included a prescription for
excitation of planetesimals by density fluctuations. They generally find (consistent with \citealt*{IGM08}, \citealt*{NG10}, \citealt*{GNT11}, \citetalias{GNT12}, and our findings) that under ideal MRI conditions (high $\alpha$ and $\gamma$) planetesimals do not accrete, but fragment. To investigate positive outcomes these studies have artificially reduced $\gamma$, motivated by a lower $\alpha$ value that may be applicable in dead zones. However, equations like (51) no longer apply in this limit.

A physically motivated way to obtain the effective $\gamma$ would follow the prescriptions outlined in this work, which are fully consistent with the detailed MHD calculations performed by \citetalias{GNT12}, but expands it to 
{ more general }conditions. An example of such an application will be presented in \citetalias{OO13b}. In closing, we also want to emphasize that the stirring conditions during the planet formation epoch do not need to be time-independent. For example, planetesimal erosion or fragmentation may
inject a large number of small grains, which potentially enlarges the
dead zone. Similarly, the net vertical magnetic flux $\brac{B_z}$ 
may evolve over time due to disk accretion \citep{RL08} 
and/or turbulent magnetic diffusion \citep{LPP94}.

\section{Summary}\label{sec:summary}

In this study, we have presented a recipe for turbulent stirring 
of planetesimals in MRI-driven turbulence.
From order-of-magnitude estimates, we have derived 
scaling relations that link the turbulent stirring rates to
the amplitude of the density fluctuations and other relevant disk parameters (Section~\ref{sec:estimate}).
The scalings do not rely on a specific choice of disk parameters, 
and hence allow to generalize the results of numerical simulations 
to wider parameter spaces. 
Our model also accounts for the effects of the shearing-out of density waves 
on the resulting stirring rates in the presence of a dead zone.
We have tested the predicted scalings using the published data 
of MHD simulations by \citetalias{GNT12} (Section~\ref{sec:calibration}).
We have shown that our scaling relations successfully explain the observed data
if we fix order-of-unity uncertainties within the relations 
(Equations~\eqref{eq:Fphi_final}, \eqref{eq:Da_final}, and \eqref{eq:De_final}).

We also have updated the saturation predictor for the density fluctuation amplitude 
in MRI-driven turbulence proposed by \citetalias{OH11} (Section~\ref{sec:predictor}).
We find that the \citetalias{OH11} predictor overestimates 
the amplitude when the MRI-active upper layer is significantly suppressed 
from above by the strongly magnetized, MRI-stable atmosphere.
To account for this effect, we have constructed a correction function, which we call the saturation limiter (Equation~\eqref{eq:L}),
on the basis of a layered MRI-turbulent disk model by \citetalias{OH11}.  
The updated predictor function (Equation~\eqref{eq:drho_new}) 
successfully reproduces the saturated amplitude of the density fluctuations 
observed in the \citetalias{GNT12} simulations. 

Combinations of the scaling relations and saturation predictor 
(Equations~\eqref{eq:Da_pred} and \eqref{eq:De_pred}) 
enable us to know how the turbulent stirring rate of planetesimals generally depends 
on disk parameters such as the gas column density, distance from the central star,
vertical resistivity distribution, and net vertical magnetic flux.
An example of such application will be presented in \citetalias{OO13b},
where we examine if runaway growth of planetesimals is possible in turbulent disks.
{ However, given that the simulations our recipe is based on only cover a limited range of the parameter space, it is yet to be warranted if our recipe applies to arbitrary disk conditions. We encourage further testing of its general validity.
}

Finally, we comment that our recipe only takes into account Ohmic diffusion 
and neglects other non-ideal MHD effects. 
Recently, \citet{BS13b} have performed local stratified MHD simulations 
with both Ohmic resistivity and ambipolar diffusion and 
concluded that ambipolar diffusion can dramatically
suppress the turbulent motion of the disk gas.
The effect of Hall diffusion is yet to be examined with stratified simulations, 
but \citet{WS12} suggest that Hall diffusion can increase or decrease 
the vertical extent of the MRI-active layers by an order of magnitude in mass.
Inclusion of these non-Ohmic effects will be an interesting extension of this work.

\acknowledgments
We thank Shigenobu Hirose for sharing with us the results of his numerical simulations. 
We are also grateful to Hiroshi Kobayashi, Takayuki Muto, Oliver Gressel, Neal Turner, 
Jeff Cuzzi, Xuening Bai, Hidekazu Tanaka, Takeru Suzuki, Shu-ichiro Inutsuka,
and the anonymous referee for useful comments.
S.O.~was supported by the Grant-in-Aid for JSPS Fellows ($22\cdot 7006$) from MEXT of Japan.
C.W.O.~acknowledges support for this work by NASA through Hubble Fellowship grant \#HST-HF-51294.01-A awarded by the Space Telescope Science Institute, which is operated by the Association of Universities for Research in Astronomy, Inc., for NASA, under contract NAS 5-26555.

\appendix
\section{Verifying Auxiliary Scaling Laws}\label{sec:OH11}
We have derived the new predictor function $[\brac{\delta\rho^2}_\md]_{\rm new}$ 
assuming two auxiliary scaling relations, Equations~\eqref{eq:dBz2_ideal} and \eqref{eq:equip}.
In this appendix, we verify these relations with the simulation data provided by \citetalias{OH11}.

\begin{figure}[t]
\epsscale{1}
\plotone{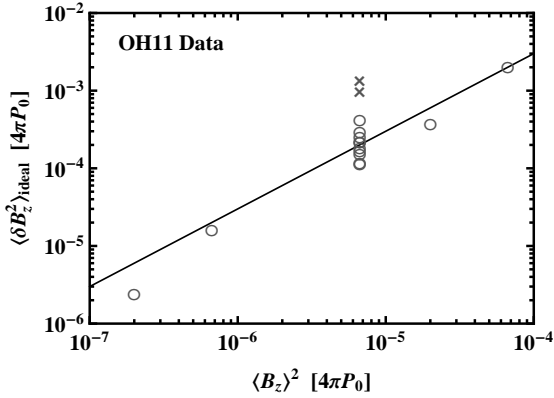}
\caption{Mean squared amplitude of the $B_z$ fluctuations measured at $z = H_\ideal$, 
$\brac{\delta B_z^2}_\ideal$, vs.~the squared net flux strength $\brac{B_z}^2$ 
for all \citetalias{OH11} simulations.
The open circles show the data for runs with $H_{\rm res,0} > 0$ while the cross symbols are 
for runs with $H_{\rm res,0} = 0$.
The line indicates $\brac{\delta B_z^2} = 30 \brac{B_z}_\ideal^2$ (see Equation~\eqref{eq:dBz2_ideal}).
}
\label{fig:A1}
\end{figure}
Figure~\ref{fig:A1} shows $\brac{\delta B_z^2}_\ideal$ versus $\brac{B_z}^2$ observed 
in the \citetalias{OH11} simulations.
Each data point corresponds to a run with a different resistivity profile.
The open circles show the data for runs with $H_{\rm res,0} > 0$ 
while the cross symbols are for runs with $H_{\rm res,0} = 0$.
We see that the data for $H_{\rm res,0} > 0$ are well explained 
by Equation~\eqref{eq:dBz2_ideal} with an accuracy of factor 2.
This is not the case for $H_{\rm res,0} = 0$, for which the high activity of MRI 
near the midplane influences the activity at high altitudes.
However, this fact does not invalidate the use of Equation~\eqref{eq:dBz2_ideal} in deriving 
the saturation limiter ${\cal L}$ because the limiter operates only when a large dead zone is present.

\begin{figure}[t]
\epsscale{1}
\plotone{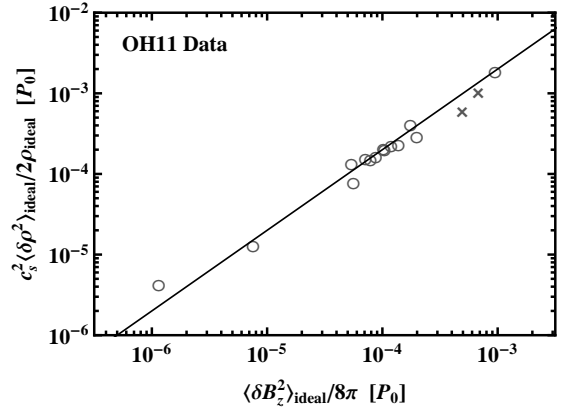}
\caption{Internal energy density of the $\rho$ fluctuations at $z=H_\ideal$, 
$c_s^2 \brac{\delta\rho^2}_\ideal/2\rho_\ideal$,
vs.~the energy density of the $B_z$ fluctuations at the same height, $\brac{\delta B_z^2}_\ideal/8\pi$, 
for \citetalias{OH11} simulations.
The open circles show the data for runs with $H_{\rm res,0} > 0 $ while the cross symbols for  runs with $H_{\rm res,0} = 0$.
The line indicates $c_s^2 \brac{\delta\rho^2}_\ideal/2\rho_\ideal = 2\brac{\delta B_z^2}_\ideal/8\pi$.
}
\label{fig:A2}
\end{figure}
Shown in Figure~\ref{fig:A2} is $c_s^2 \brac{\delta\rho^2}_\ideal/2\rho_\ideal$ 
versus $\brac{\delta B_z^2}_\ideal/8\pi$ observed in the \citetalias{OH11} simulations.
The data are well fitted by $c_s^2 \brac{\delta\rho^2}_\ideal/2\rho_\ideal = 2\brac{\delta B_z^2}_\ideal/8\pi$, which supports Equation~\eqref{eq:equip}.
A similar scaling is also observed in local unstratified simulations \citep{SITS04}.




\begin{thebibliography}{}

\bibitem[Bai(2011)]{B11a}
Bai, X.-N. 2011, \apj, 739, 50

\bibitem[Bai \& Stone(2013a)]{BS13a}
Bai, X.-N., \& Stone, J. M. 2013a, \apj, 767, 30

\bibitem[Bai \& Stone(2013b)]{BS13b}
Bai, X.-N., \& Stone, J. M. 2013b, \apj, in press (arXiv:1301.0318)

\bibitem[Balbus \& Hawley(1991)]{BH91}
Balbus, S. A., \& Hawley, J. F. 1991, \apj, 376, 214

\bibitem[Baruteau \& Lin(2010)]{BL10}
Baruteau C., \& Lin D. N. C. 2010, \apj, 709, 759

\bibitem[Carballido et al.(2005)]{CSP05}
Carballido, A., Stone, J. M., \& Pringle, J. E. 2005, \mnras, 358, 1055

\bibitem[Cuzzi et al.(2001)]{C+01}
Cuzzi, J. N., Hogan, R. C., Paque, J. M., \& Dobrovolskis, A. R. 2001, \apj, 546, 496

\bibitem[Davis et al.(2010)]{DSP10}
Davis, S. W., Stone, J. M., \& Pessah, M. E. 2010, \apj, 713, 52

\bibitem[Dzyurkevich et al.(2013)]{DTHK13}
Dzyurkevich, N., Turner, N. J., Henning, Th., \& Kley, W. 2013, \apj, 
765, 114

\bibitem[Fromang \& Papaloizou(2006)]{FP06}
Fromang, S., \& Papaloizou, J. 2006, \aap, 452, 751

\bibitem[Gammie(1996)]{G96}
Gammie, C. F. 1996, \apj, 457, 355

\bibitem[Goldreich \& Lynden-Bell(1965)]{GL65}
Goldreich, P., \& Lynden-Bell, D. 1965, \mnras, 130, 125 

\bibitem[Goldreich \& Ward(1973)]{GW73}
Goldreich, P., \& Ward, W. R. 1973, \apj, 183, 1051

\bibitem[Gressel et al.(2011)]{GNT11}
Gressel, O., Nelson, R. P., \& Turner, N. J. 2011, \mnras, 415, 3291

\bibitem[Gressel et al.(2012)]{GNT12}
Gressel, O., Nelson, R. P., \& Turner, N. J. 2012, \mnras, 422, 1240 (GNT12)

\bibitem[Guan et al.(2009)]{GGSJ09}
Guan, X., Gammie, C. F., Simon, J. B., \& Johnson, B. M. 2009, ApJ, 694, 1010

\bibitem[Hawley et al.(1995)]{HGB95}
Hawley, J. F., Gammie, C. F., \& Balbus, S. A. 1995, \apj, 440, 742

\bibitem[Heinemann \& Papaloizou(2009a)]{HP09a}
Heinemann, T., \& Papaloizou, J. C. B. 2009a, \mnras, 397, 52

\bibitem[Heinemann \& Papaloizou(2009b)]{HP09b}
Heinemann, T., \& Papaloizou, J. C. B. 2009b, \mnras, 397, 64

\bibitem[Ida et al.(2008)]{IGM08}
Ida, S., Guillot, T., \& Morbidelli, A. 2008, \apj, 686, 1292

\bibitem[Ilgner \& Nelson(2006)]{IN06a}  
Ilgner, M., \& Nelson, R. P. 2006, \aap, 445, 205

\bibitem[Jin(1996)]{J96}
Jin, L. 1996, \apj, 457, 798

\bibitem[Johansen et al.(2008)]{J+08}
Johansen, A., Brauer, F., Dullemond, C., Klahr, H., \& Henning, T. 2008, \aap, 486, 597

\bibitem[Johansen et al.(2006)]{JKM06}
Johansen, A., Klahr, H., \& Mee, A. J. 2006, \mnras, 370, L71

\bibitem[Johansen et al.(2007)]{J+07}
Johansen, A., Oishi, J. S., Mac Low, M.-M., et al. 2007, Nature, 448, 1022

\bibitem[Johnson et al.(2006)]{JGM06}
Johnson, E. T., Goodman, J., \& Menou, K. 2006, \apj, 647, 1413

\bibitem[Kokubo \& Ida(1996)]{KI96}
Kokubo, E., \& Ida, S. 1996, Icarus, 123, 180

\bibitem[Laughlin et al.(2004)]{LSA04}
Laughlin, G., Steinacker, A., \&  Adams, F. C. 2004, \apj, 608, 489

\bibitem[Lubow et al.(1994)]{LPP94}
Lubow, S. H., Papaloizou, J. C. B., \& Pringle, J. E. 1994, MNRAS, 267, 235

\bibitem[Meschiari(2012)]{M12}
Meschiari, S. 2012, \apj, 752, 71 

\bibitem[Mohanty et al.(2013)]{MET13}
Mohanty, S., Ercolano, B, \& Turner, N. J. 2013, \apj, 764, 65

\bibitem[Morbidelli et al.(2009)]{MBNL09}
Morbidelli, A., Bottke, W. F., Nesvorny, D., \& Levison, H. F. 2009,
Icarus, 204, 558 

\bibitem[Nelson(2005)]{N05}
Nelson, R. P. 2005, \aap, 443, 1067

\bibitem[Nelson \& Gressel(2010)]{NG10}
Nelson, R. P., \& Gressel, O. 2010, \mnras, 409, 639

\bibitem[Nelson \& Papaloizou(2004)]{NP04}
Nelson, R. P., \& Papaloizou, J. C. B. 2004, \mnras, 350, 849

\bibitem[Ogihara et al.(2007)]{OIM07}
Ogihara, M., Ida, S., \& Morbidelli, A. 2007, Icarus, 188, 522

\bibitem[Oishi et al.(2007)]{OMM07}
Oishi, J. S., Mac Low, M.-M., \& Menou, K. 2007, \apj, 670, 805

\bibitem[Okuzumi(2009)]{O09}
Okuzumi, S. 2009, \apj, 698, 1122

\bibitem[Okuzumi \& Hirose(2011)]{OH11} 
Okuzumi, S., \& Hirose, S. 2011, \apj, 742, 65 (OH11)

\bibitem[Okuzumi \& Hirose(2012)]{OH12} 
Okuzumi, S., \& Hirose, S. 2012, \apjl, 743, L8

\bibitem[Okuzumi et al.(2012)]{OTKW12}
Okuzumi, S., Tanaka, H., Kobayashi, H., \& Wada, K. 2012, \apj, 752, 106

\bibitem[Ormel \& Cuzzi(2007)]{OC07}
Ormel, C. W., \& Cuzzi, J. N. 2007, \aap, 466, 413

\bibitem[Ormel \& Okuzumi(2013)]{OO13b}
Ormel, C. W., \& Okuzumi, S. 2013, \apj, in press (Paper II)

\bibitem[Ormel et al.(2010)]{ODS10b}
Ormel, C. W., Dullemond, C. P., \& Spaans, M. 2010, Icarus, 210, 507

\bibitem[Pan et al.(2011)]{P+11}	
Pan, L., Padoan, P., Scalo, J., Kritsuk, A. G., \& Norman, M. L. 2011, \apj, 740, 6

\bibitem[Perez-Becker \& Chiang(2011a)]{PC11a}
Perez-Becker, D., \& Chiang, E. 2011a, \apj, 727, 2

\bibitem[Perez-Becker \& Chiang(2011b)]{PC11b}
Perez-Becker, D., \& Chiang, E. 2011b, \apj, 735, 8

\bibitem[Rein(2012)]{R12}
Rein, H. 2012, \mnras, L526 

\bibitem[Rein \& Papaloizou(2009)]{RP09}
Rein, H., \& Papaloizou, J. C. B. 2009, \aap, 497, 595
 
\bibitem[Rothstein \& Lovelace(2008)]{RL08}
Rothstein, D. M., \& Lovelace, R. V. E. 2008, \apj, 677, 1221
 
\bibitem[Sano et al.(2004)]{SITS04}
Sano, T., Inutsuka, S., Turner, N. J., \& Stone, J. M. 2004, \apj,  605, 321

\bibitem[Sano \& Miyama(1999)]{SM99}
Sano, T., \& Miyama, S. M. 1999, \apj,  515, 776

\bibitem[Sano et al.(2000)]{SMUN00}
Sano, T., Miyama, S. M., Umebayashi, T., \& Nakano, T.  2000, \apj,  543, 486

\bibitem[Suzuki et al.(2010)]{SMI10}
Suzuki, T. K., Muto, T., \& Inutsuka, S. 2010, \apj, 718, 1289

\bibitem[Turner \& Sano(2008)]{TS08}
Turner, N. J., \&  Sano, T. 2008, \apj, 679, L131

\bibitem[Turner et al.(2006)]{TWBY06}
Turner, N. J., Willacy, K., Bryden, G., \& Yorke, H. W. 2006, \apj, 639, 1218

\bibitem[V\"{o}lk et al.(1980)]{V+80}
V\"{o}lk, H. J., Jones, F. C., Morfill, G. E., \& R\"{o}ser S. 1980, \aap, 85, 316

\bibitem[Wardle \& Salmeron(2012)]{WS12}
Wardle, M.. \& Salmeron, R. 2012, \mnras, 422, 2737

\bibitem[Weidenschilling(2011)]{W11}
Weidenschilling, S. J. 2011, Icarus, 214, 671

\bibitem[Wetherill \& Stewart(1989)]{WS89}
Wetherill, G. W., \& Stewart, G. R. 1989, Icarus, 77, 330 

\bibitem[Windmark et al.(2012)]{W+12a} 
Windmark, F., Birnstiel, T., G{\"u}ttler, C., et al. 2012, \aap, 540, A73 

\bibitem[Yang et al.(2009)]{YMM09}
Yang, C.-C., Mac Low, M.-M., \& Menou, K. 2009, \apj, 707, 1233

\bibitem[Yang et al.(2012)]{YMM12}
Yang, C.-C., Mac Low, M.-M., \& Menou, K. 2012, \apj, 748, 79

\bibitem[Youdin(2011)]{Y11}
Youdin, A. N. 2011, \apj, 731, 99

\bibitem[Youdin \& Goodman(2005)]{YG05}
Youdin, A. N., \& Goodman, J. 2005, \apj, 620, 459

\end{thebibliography}
\end{document}